\documentclass[pdflatex,sn-basic]{sn-jnl}

\usepackage[T1]{fontenc}
\usepackage[utf8]{inputenc}
\usepackage{graphicx}%
\usepackage{multirow}%
\usepackage{amsmath,amssymb,amsfonts}%
\usepackage{amsthm}%
\usepackage{mathrsfs}%
\usepackage[title]{appendix}%
\usepackage{xcolor}%
\usepackage{textcomp}%
\usepackage{manyfoot}%
\usepackage{booktabs}%
\usepackage{algorithm}%
\usepackage{algorithmicx}%
\usepackage{algpseudocode}%
\usepackage{listings}%
\usepackage{dirtree,array}
\usepackage{enumitem}
\usepackage{soul} 
\usepackage{url}
\usepackage{tikz}
\usepackage{xcolor}
\usepackage{tcolorbox}
\tcbuselibrary{listings,skins,breakable}
\usepackage{listings}
\usetikzlibrary{shapes.misc, shapes,arrows.meta,positioning,calc,fit,backgrounds,decorations.pathreplacing}
\newcommand{\mytimes}{ \tikz[baseline=-.55ex] \node [inner sep=0pt,cross out,draw,line width=1pt,minimum size=1ex] (a) {};}

\newcommand{\emsejon}[1]{\textcolor{black}{#1}}

\newcommand{\emsemaksym}[1]{\textcolor{black}{#1}}

\newcommand{\emseparis}[1]{\textcolor{black}{#1}}

\definecolor{systembg}{RGB}{230,243,255}      
\definecolor{systemframe}{RGB}{70,130,180}    
\definecolor{userbg}{RGB}{232,245,233}        
\definecolor{userframe}{RGB}{76,175,80}       
\definecolor{outputbg}{RGB}{245,245,245}      
\definecolor{outputframe}{RGB}{120,120,120}   
\definecolor{highlightbg}{RGB}{255,248,225}   
\definecolor{highlightframe}{RGB}{255,160,0}  

\newtcolorbox{systemmsg}[1][]{
  colback=systembg,
  colframe=systemframe,
  fonttitle=\bfseries\small,
  title={\textsc{System Message}},
  boxrule=0.5pt,
  left=6pt, right=6pt, top=4pt, bottom=4pt,
  fontupper=\small,
  breakable,
  #1
}

\newtcolorbox{usermsg}[1][]{
  colback=userbg,
  colframe=userframe,
  fonttitle=\bfseries\small,
  title={\textsc{User Message (excerpt)}},
  boxrule=0.5pt,
  left=6pt, right=6pt, top=4pt, bottom=4pt,
  fontupper=\small,
  breakable,
  #1
}

\newtcolorbox{outputschema}[1][]{
  colback=outputbg,
  colframe=outputframe,
  fonttitle=\bfseries\small,
  title={\textsc{Output Schema}},
  boxrule=0.5pt,
  left=6pt, right=6pt, top=4pt, bottom=4pt,
  fontupper=\small\ttfamily,
  breakable,
  #1
}

\newtcolorbox{keyhighlight}[1][]{
  colback=highlightbg,
  colframe=highlightframe,
  fonttitle=\bfseries\small,
  boxrule=0.8pt,
  left=6pt, right=6pt, top=4pt, bottom=4pt,
  fontupper=\small,
  breakable,
  #1
}

\theoremstyle{thmstyleone}%
\theoremstyle{thmstyletwo}%

\theoremstyle{thmstylethree}%

\begin{document}

\title[Article Title]{Evaluating Non-English Developer Support in Machine Learning for Software Engineering}


\author*[1]{\fnm{Jonathan} \sur{Katzy}}\email{J.B.Katzy@TUDelft.nl}

\author[1]{\fnm{Yongcheng} \sur{Huang}}\email{Y.Huang-51@Student.TUDelft.nl}

\author[1]{\fnm{Gopal-Raj} \sur{Panchu}}\email{G.G.S.Panchu@Student.TUDelft.nl}

\author[1]{\fnm{Maksym} \sur{Ziemlewski}}\email{M.Ziemlewski@Student.TUDelft.nl}

\author[1]{\fnm{Paris} \sur{Loizides}}\email{P.Loizides@Student.TUDelft.nl}

\author[1]{\fnm{Sander} \sur{Vermeulen}}\email{S.R.Vermeulen@Student.TUDelft.nl}

\author[1]{\fnm{Arie} \sur{van Deursen}}\email{Arie.vanDeursen@TUDelft.nl}

\author[1]{\fnm{Maliheh} \sur{Izadi}}\email{M.Izadi@TUDelft.nl}


\affil*[1]{\orgdiv{Software Engineering Research Group}, \orgname{Delft University of Technology}, \country{The Netherlands}}




\abstract{Large Language Models are increasingly used in software engineering, but both code generation and its evaluation remain predominantly English-centric. This leaves a major gap in our understanding of how well current tools support multilingual development, where code contains non-English natural language. In this paper, we investigate non-English code comment generation and the reliability of current methods for evaluating such outputs. We evaluate five code LLMs (CodeGemma, CodeLlama, CodeQwen1.5, GraniteCode, and StarCoder2) across five natural languages: Dutch, English, Greek, Polish and Chinese. We further conduct an open-coding study of 12,500 generated comments, from which we derive a publicly released human-annotated dataset and a taxonomy of 26 error types. We use these human annotations, to evaluate the performance of neural metrics, and LLM-as-a-judge pipelines.

Our findings show that generative performance deteriorates substantially outside English, with linguistic errors increasing by up to 15.1$\times$, alongside more frequent incoherent generations and a moderate rise in semantic errors. \emsejon{More critically, we show that detecting errors in non-English comments underperforms. Across classical overlap-based metrics, off-the-shelf neural metrics, extended neural metrics using newer multilingual, language-specific, and code-specific models, and LLM-as-a-judge pipelines, no automatic approach consistently provides reliable and consistent assessment. Neural metrics often fail to distinguish correct comments from incorrect outputs or even random noise, and tend to overestimate quality in non-English settings. This is not resolved by adding context or using specialized scoring models. LLM-as-a-judge methods achieve the highest agreement with human annotations but fail to reliably capture important language-related and semantic errors. Overall, our results show that evaluation and generation are key barriers to progress for multilingual tooling, and that human judgment remains indispensable in this setting.}}




\maketitle
\section{Introduction}
The adoption of Large Language Models (LLMs) for code has led to numerous benefits. When integrated into the software development process, studies reveal that LLMs boost developer productivity~\citep{ziegler2022productivityassessmentneuralcode} and facilitate various software engineering tasks~\citep{jiang2024survey,izadi2022codefill}. LLMs have also been shown to support programming education~\citep{boguslawski2025programming}, and assist developers of various skill levels~\citep{nam2024using}.

Although code models can enhance both learning and development, most research and development in this area focus on \textit{English} contexts. This emphasis limits the adoption of code models for other languages, excluding non-English speakers from the benefits of LLMs in education~\citep{10.1145/3657604.3662036}, access to learning in their preferred language~\citep{language_sweden, 10.1145/2729094.2742618}, and integration of code models into workflows that often involve substantial non-English code~\citep{hu2022practitioners, 7332491}. Beyond adoption challenges, non-English code is sometimes considered a ``data smell'' to be avoided when training code models~\citep{vitale2024catalog} leading to progressively reduced representation. All these issues combine in a lack of scientific literature studying how well non-English developers are supported in the machine learning for software engineering (ML4SE) ecosystem.

\emsejon{By non-English code, we refer to the presence of a language other than English in either the identifiers, literals, or comments. This is often a mixture of English and another language. For software engineering tasks assisted by an LLM, the presence of these artifacts is crucial and can improve performance on tasks like automated bug-fixing by up to threefold ~\citep{vitale2026impact}. Traditionally, non-English code was not commonly used in open source projects; however, closed-source proprietary codebases often contain other languages~\citep{7332491}. Recently, a shift has been observed in the open-source landscape, and multilingual participation in open-source repositories is steadily increasing across code comments, strings, and documentation~\citep{bhuiyan2026write}.}

\emsejon{To situate this problem, we consider the machine learning for software engineering (ML4SE) ecosystem as comprising both generative and evaluation techniques. Generative techniques are used for tasks such as code generation, summarization, bug localization and fixing, and comment generation. Their outputs must then be assessed, which is handled by evaluation techniques. In this investigation we evaluate the ability of LLMs to generate non-English comments, and then compare human labels to quality scores generated by a range of automatic scoring metrics. We then evaluate the metrics themselves based on their alignment with humans, reliability, robustness, and coverage of errors detected.}

For the generative component, we focus on code comment generation as a representative task that requires both code understanding and language proficiency. We evaluate five state-of-the-art code LLMs (CodeGemma~\citep{codegemmateam2024}, CodeLlama~\citep{roziere2023code}, CodeQwen1.5~\citep{codeqwen1.5}, GraniteCode~\citep{mishra2024granitecodemodelsfamily}, and StarCoder2~\citep{lozhkov2024starcoder2stackv2}) across five natural languages (\textit{English}, \textit{Dutch}, \textit{Chinese}, \textit{Greek}, and \textit{Polish}).

For the qualitative evaluation, we conduct an open-coding study to identify common error types in model-generated comments and assign correctness labels (\texttt{correct}, \texttt{partially correct}, \texttt{incorrect}). This analysis characterizes key failure modes and provides insight into challenges related to trust~\citep{Witchel2022},
education~\citep{nishanthi2020understanding}, and adoption in non-English workflows~\citep{hu2022practitioners}, forming the basis for the following research questions:
\begin{itemize}
    \item[\textbf{RQ1.1}] What types of errors do LLMs make when generating comments in different languages?
    \item[\textbf{RQ1.2}] How accurately do LLMs generate comments for non-English code?
\end{itemize}

Using these human annotations, we evaluate the reliability of off-the-shelf metrics for code comment generation, focusing on metrics that are readily available through standard libraries and widely used in practice. We examine
their alignment with human judgment, their ability to distinguish correct from incorrect outputs, and their robustness to perturbations, addressing the following research questions:
\begin{itemize}
    \item[\textbf{RQ2.1}] Do off-the-shelf evaluation metrics give a trustworthy evaluation of LLM generated code comments?
    \item[\textbf{RQ2.2}] Do off-the-shelf metrics effectively differentiate between correct and incorrect predictions?
    \item[\textbf{RQ2.3}] Are off-the-shelf metrics robust to perturbations in the input?
\end{itemize}

\emsejon{Building on these findings, we extend neural evaluation metrics such as BERTScore and BARTScore by incorporating more recent, language-specific, and code-specific scoring models. We also vary the amount of contextual information provided to the models. We evaluate how these extensions affect alignment with human judgment, the ability to leverage contextual information, and robustness to perturbations, addressing the following research questions:}
\begin{itemize}
    \item[\textbf{RQ3.1}] \emsejon{Which neural evaluation models best align with human judgment?}
    \item[\textbf{RQ3.2}] \emsejon{Do neural metrics capture the nuances present in the surrounding context?}
    \item[\textbf{RQ3.3}] \emsejon{Are neural metrics robust to perturbations in the input?}
\end{itemize}

\emsejon{Finally, we investigate whether LLM-as-a-judge pipelines can reliably replace human evaluation in non-English settings. We examine their limitations, their alignment with human judgment, and their ability to detect specific error types, addressing the following research questions:}
\begin{itemize}
    \item[\textbf{RQ4.1}] \emsejon{What are the limitations of LLM-as-a-judge approaches in non-English contexts?}
    \item[\textbf{RQ4.2}] \emsejon{Do LLM-as-a-judge setups align with human judgment when identifying correct comments?}
    \item[\textbf{RQ4.3}] \emsejon{Do LLM-as-a-judge setups effectively detect individual error types?}
\end{itemize}
\emsejon{
Our results reveal significant limitations in current ML4SE support for non-English code. From our qualitative analysis, we derive a taxonomy of 26 error types and observe that linguistic errors increase substantially in non-English contexts (up to $15.1\times$), alongside a rise in incoherent generations, while semantic errors increase more moderately. For evaluation, we find that both classical and neural off-the-shelf metrics struggle to reliably distinguish correct outputs from incorrect ones. In particular, neural metrics exhibit a limited sensitivity to random noise and tend to assign overly high scores, especially in non-English settings. This behavior contrasts with classical overlap-based metrics and highlights fundamental differences in how these approaches respond to noisy or incorrect inputs. Extending neural metrics with more recent, language-specific, and code-specific models does not resolve these issues, and incorporating additional context does not consistently improve performance. Finally, LLM-as-a-judge approaches exhibit instability across languages, tend to be overly critical, and fail to reliably detect key error types, especially those related to language proficiency and semantic correctness. These findings highlight substantial challenges for the reliable adoption of code LLMs in non-English workflows.}

This paper makes the following contributions:
\begin{itemize}
    \item We present a comprehensive empirical study of LLM performance on non-English code comment generation across five natural languages, highlighting substantial differences in error patterns compared to English.
    \item \emsejon{We introduce a taxonomy of 26 error types and publicly release a human-annotated dataset of 12,500 code comment samples spanning five LLMs and five languages, enabling fine-grained analysis of LLM behavior in multilingual code settings, and per model results of all settings in the supporting material.}
    \item We evaluate common off-the-shelf evaluation metrics and show that neural metrics struggle to distinguish correct outputs from incorrect or random noise.
    \item \emsejon{We extend neural evaluation metrics with more context and show that this does not aid them in separating correct from incorrect predictions.}
    \item\emsejon{ We evaluate LLM-as-a-judge pipelines in non-English settings and identify key limitations, including instability across languages and difficulty in detecting language-related and semantic errors.}
\end{itemize}

\section{Related Work}
\paragraph{Multilingual LLMs in NLP}
Studies on LLMs in NLP settings have shown that they perform poorly in non-English and resource-poor natural languages. The performance of several models investigated in Indonesian and African languages continues to be poor, producing responses of lower quality compared to other high-resource languages such as English~\citep{koto2023largelanguagemodelspass, ojo2024goodlargelanguagemodels}. Further studies showed lower performance for non-English queries in chat models, resulting in less accurate and lower quality responses compared to English queries, as well as issues arising with regard to cultural nuances when dealing with non-English queries~\citep{zhang2023donttrustchatgptquestion}.
Such findings are consistent with research suggesting a strong bias in multilingual settings when generating code based on non-English instructions. In the case of Code LLaMA, a performance drop $37.8\%$ (Pass@1 metric) was observed when code generation was prompted in Chinese rather than English~\citep{wang2024exploringmultilingualbiaslarge}. This performance gap emphasizes the underlying bias in LLMs trained predominantly on English-language datasets. 

\paragraph{Comment/Documentation Generation}
Automatic comment generation is a common task that has been shown to help developers understand code~\citep{khan2022automaticcodedocumentationgeneration}. 
In this task, language models are used to generate a natural language description of a piece of code. Although it has been shown to improve developer productivity, most studies focus primarily on the performance of models in English code. 
Aside from evaluating models on comment generation, research has shown that for automated comment generation in general; practitioners expect automatically generated comments to be fluent and grammatically correct (82\%), be concise (88\%), and accurate (88\%)~\citep{hu2022practitioners} in order to include them in their workflows.

\paragraph{Qualitative Analyses of LLM Outputs}
Previous works on translation and specializations on automatic translations have led to the development of Taxonomies like SCATE~\citep{8544345} and evaluation frameworks such as MQM~\citep{mariana2014multidimensional}, used for natural language tasks. These provide a basis for error classification; however, they do not capture all intricacies of code related tasks. Further qualitative studies of code generation errors have identified common bugs in LLM-generated code~\citep{tambon2024bugslargelanguagemodels, code4me} and given important insights into user preference for model invocations and acceptance rates~\citep{code4me}, however, they make the assumption that code is always written in English, and they focus primarily on the correctness of a program, rather than specific errors in the language.

\paragraph{Non-English Code}
Despite the global prevalence of code written in different languages, non-English code remains significantly under-researched. To our knowledge, only one study has examined the distribution of non-English language in code repositories~\citep{7332491}. They show that there is a significant increase in the use of German in both comments and identifier names when looking at industry projects, compared to open source projects, which showed to have almost no non-English text present. \emsejon{However, a recent large-scale study counters this trend, revealing that multilingual participation in open-source code is steadily increasing, particularly in Korean and Chinese, across code comments, strings, and documentation~\citep{bhuiyan2026write}. The authors show that non-English speaking communities are among the fastest-growing demographics on GitHub, which further highlights the need for multilingual LLM evaluation.} Datasets that are intended to be used for the training of LLMs do mention the presence of other languages in the code, however, do not give any more specific information~\citep{lozhkov2024starcoder2stackv2}.

For code LLMs, most research focuses on English contexts; even researchers who are not native English speakers frame their studies within an English paradigm~\citep{LpezNavarro2015WhyDI}, leading to an under-representation of non-English scenarios. GitHub reports show that a significant number of non-English speakers contribute code to the platform each year, highlighting the importance of studying code generation in non-English contexts~\citep{octoverse2022}.

\paragraph{Metrics Reliability}
The ability to evaluate the quality of text generated by LLMs is important to fairly evaluate and compare different models. However, there are some challenges that arise when attempting to programmatically evaluate natural language generation. 
First, many early methods for evaluating textual generations come from the translation domain, where a generation can be compared to multiple reference texts in order to score them appropriately~\citep{papineni2002bleu, lin2004rouge, banerjee2005meteor}. However, these models mainly look at an overlap of letters in the target generation with the reference generations. This creates issues when comparing the correctness of the meaning behind the text, which has been proposed to be solved with embedding based metrics~\citep{zhang2019bertscore, zhou2023codebertscore} or model based metrics~\citep{yuan2021bartscore}.
This has led to the development of frameworks that can be used to compare metrics with each other~\citep{xiao2023evaluating} as well as studies measuring the correlation between different metrics~\citep{Fabbri2020SummEvalRS} and their resilience to simple perturbations~\citep{sai-etal-2021-perturbation}. These studies give an overview of what is to be expected from a metric to be a good judge of generated text. We summarize the requirements as follows: A metric must differentiate correct from incorrect samples, agree with human expert expectations, and be able to detect simple perturbations.

\paragraph{LLM-as-a-Judge}
\emsejon{
In LLM as a Judge setting, an LLM evaluates the output of another model according to a specified rubric, whether assigning scores or selecting the better response in pairwise comparisons. This approach enables scalable evaluation and reduces the reliance on human annotation~\citep{chehbouni2025neither, li2024llms}. 
Recent surveys provide a broad overview of this paradigm and its adoption across NLP and software engineering~\citep{gu2024surveyllmjudge, li2025generation2judgment}. A literature review focused on 42 SE studies notes that code comment and documentation quality evaluation remains underexplored within this line of work~\citep{he2025code2courtroom}.
In the software engineering domain, this paradigm has been applied to code generation and code summarization, where LLM judges achieve moderate to substantial agreement with human annotators, though even the best-performing models frequently misjudge output quality~\citep{crupi2025effectiveness}.
It has also been shown that for certain SE annotation tasks, including code summarization quality, one human rater can be replaced by an LLM, with model-model agreement being a useful predictor of whether a task is suitable for this~\citep{ahmed2025llmannotation}. However, the effectiveness of LLM judges is task-dependent. Output-based methods show near-human alignment for code translation but are weaker for code summarization~\citep{wang2025llmjudgeSE}. This indicates that tasks involving natural language quality, such as code comment evaluation, are still difficult for automated judges.}

\emsejon{Previous work shows that LLM judges exhibit biases in evaluations, such as the self-bias of assigning higher scores to outputs generated by models from the same family~\citep{spiliopoulou2025play}. Strong sensitivity to prompt phrasing is also demonstrated~\citep{chen2024humans}, along with positional bias, where changing the order of candidate responses in the prompt influences their quality ranking~\citep{wang2024large}.}

\emsejon{In addition, LLMs have been shown to exhibit sycophantic behavior, with responses aligning with user statements rather than independent verification~\citep{fanous2025syceval,malmqvist2025sycophancy}. 
\emseparis{This behavior has been attributed to reinforcement learning from human feedback (RLHF) training~\citep{christiano2017deep}, where human preference data systematically favors agreeable responses over truthful ones~\citep{sharma2024sycophancy}.} 
This trend raises concerns when LLMs are used as evaluators, but despite these limitations, LLM-based evaluators are adopted in practice because they can automate evaluation workflows and reduce the cognitive load associated with manual annotation~\citep{chehbouni2025neither}.}

\emsejon{A limitation that is directly relevant to our work is the reliability of LLM judges across languages. Studies on multilingual LLM-as-a-Judge show poor cross-language consistency, with neither multilingual fine-tuning nor larger model sizes leading to reliable improvements~\citep{fu2025multilingual}. Evaluations of reward models across languages show a similar trend, where performance drops from English to non-English languages, and lower-resource languages are affected more~\citep{gureja2025mrewardbench}. To date, no work has studied multilingual LLM-as-a-Judge evaluation for code comments, where both natural language proficiency and code understanding need to be assessed. Our work addresses this.}

\emsejon{A further challenge for LLM-based evaluation concerns the cognitive demands of large label spaces. LLM performance has been shown to degrade when relevant information appears in the middle of long input contexts~\citep{liu2024lostmiddle}. More generally, decomposing complex tasks into focused sub-tasks consistently outperforms monolithic prompting~\citep{khot2023decomposed}. Together with work on structured LLM classification~\citep{lee2024improvingllmclassificationlogical}, these findings motivate our hierarchical evaluation design, which separates the error taxonomy into semantically coherent clusters to reduce the label space and cognitive load per evaluation pass.}

\section{Non-English Characteristics}
\label{non-english}
To gain an overview of the differences in the selected languages compared to English, we list the defining characteristics and how they may influence the output of a model.

\paragraph{Morphosyntactic Complexity}
Morphosyntactic features govern how words change form (morphology) and combine into sentences (syntax). The Chinese uninflected monomorphemic system eliminates grammatical marks such as tense/case~\citep{pulleyblank1995outline}, while Polish uses seven grammatical cases and conjugations with gender. Greek combines gendered nouns with flexible word order through inflectional endings, and Dutch employs gendered articles (de/het). These differences add a layer of complexity to the generation of languages compared to English. Furthermore, both Polish and Greek are flexible when it comes to following a Subject Verb Object (SVO) structure of a sentence, making the outputs context dependent~\citep{HaiderSzucsich+2022+1+39, tzanidaki1995greek}

\paragraph{Orthographic Challenges}
Orthographic systems determine how languages represent speech in writing. Greek's technical symbols (e.g., \(\alpha\), \(\beta\)) and diacritics (tonos/diaeresis) coexist with Polish's nine accented characters (ą, ó, etc.), while Chinese uses logograms instead of alphabetic writing. These add a layer of complexity to the tokenization of text and require language-specific tokens.
Furthermore, these languages face informal script variations: Greeklish (Latinized Greek), diacritic omission in Polish code comments, and Chinese character simplification. These are informal ways of writing a language, often adopted in online spaces that add complexity to the model of learning~\citep{toumazatos-etal-2024-still}. 
Finally, Greek has also been adopted as a common alphabet for use in mathematics; this adds another source of noise that could reduce a model's ability to model Greek.

\paragraph{Semantic Nuances}
Semantic features require an understanding of meaning beyond direct translation. Chinese requires contextual interpretation of idioms~\citep{10.1162/tacl_a_00572}, Dutch incorporates English loanwords through prolonged contact~\citep{VERHEIJEN2022100091}, while Polish/Greek use inflection-enabled word order flexibility to convey pragmatic emphasis. Such features demand cultural and syntactic awareness beyond direct translation equivalents, challenging LLMs' ability to preserve intended meaning during code comment generation.

\section{Overall-setup}
The basis of our investigation is a dataset of labeled code comments in 5 separate languages. First we lay out the models that are used, and then we explain the raw data collection procedure, and comment generation procedure.

\subsection{Targeted Models}
To generate code comments, we use open-weight models. This gives us full control over the generations. Ensuring that no other pre-processing steps are performed unknown to us, and allows for reproducible research, which is not guaranteed when using proprietary or hosted solutions~\citep{semmelrock2024reproducibilitymachinelearningbasedresearch}. More specifically, we focus on models within the 7B-8B parameter range, the largest number of parameters that have releases of all $5$ model families. To understand the prevalence of our target languages in the training procedure of these models, we present a brief overview.

\paragraph{CodeQwen1.5-7B}
CodeQwen1.5-7B extends Qwen1.5 with an English and Chinese dataset, supplemented by 3T tokens of code data spanning 92 programming languages, although the distribution of natural languages in this corpus is not specified~\citep{codeqwen1.5, qwen1.5}.

\paragraph{StarCoder2-7B}
StarCoder2-7B is trained on The Stack V2 dataset\footnote{\url{https://huggingface.co/datasets/bigcode/the-stack-v2}}, totaling 525.5B tokens of programming languages and documentation, with additional natural language data from Wikipedia, StackOverflow, Arxiv, Free Programming Books~\footnote{\url{https://github.com/EbookFoundation/free-programming-books}}, and OpenWebMath, covering all languages in our study~\citep{lozhkov2024starcoder2stackv2}.

\paragraph{Granite-8B-code-base}
Granite-8B-code-base is trained on a collection of programming language datasets, including GitHub Code Clean\footnote{https://huggingface.co/datasets/codeparrot/github-code-clean}, StarCoderData\footnote{https://huggingface.co/datasets/bigcode/starcoderdata}, and GitHub issues, combined with natural language datasets from StackExchange, Arxiv, and OpenWebMath. English is predominant, with non-English data removed~\citep{mishra2024granitecodemodelsfamily}.

\paragraph{CodeLlama-7B}
CodeLlama-7B is based on on Llama2, whose training corpus is 89.7\% English, with minor representation of Chinese, Dutch and Polish. CodeLlama is further trained on publicly available code, incorporating 8\% natural language data related to code and 7\% from natural language dataset, although specific origins remain undisclosed~\citep{roziere2023code, touvron2023llama}.

\paragraph{CodeGemma-7B}
CodeGemma-7B originates from the Gemma family, trained on 6T tokens inspired by Gemini, the training details of which remain undisclosed. It is fine-tuned on a dataset consisting of 80\% code and 20\% natural language, predominantly English, with no specified multi-lingual focus. However, dataset sources are also undisclosed~\citep{codegemmateam2024, gemmateam2024gemmaopenmodelsbased}.

\subsection{Data Collection and Selection}

\begin{figure}
    \centering
    \includegraphics[width=0.8\linewidth]{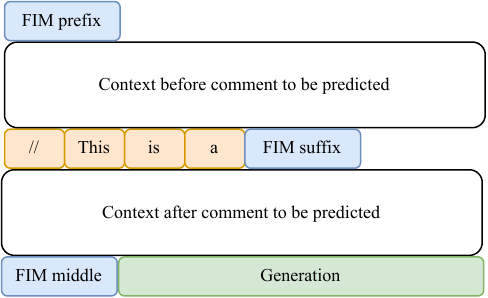}
    \caption{Example input used for inference}
    \label{fig:input_format}
\end{figure}


To gather a code dataset for each natural language, a list of the $2,500$ most common words in each language was used~\footnote{https://github.com/oprogramador/most-common-words-by-language}, based on the OpenSubtitles2016 dataset~\citep{lison-tiedemann-2016-opensubtitles2016}. For each word, the GitHub API was queried to gather 100 files containing the specific word. The collected files were then de-duplicated. 

Due to varying architectures, each model has its own maximum token length. To ensure a fair comparison between models, we filter the candidate files by their length after tokenization. We then select only files where all the context fits within the context of the smallest model (CodeGemma~\citep{codegemmateam2024}, $4,096$ tokens).
To define the inference limit, we calculated the average comment length plus three standard deviations of the ground truth of comments. We also filtered out all files where the context length + average length + $3$ standard deviations was greater than $4,096$ tokens to ensure that all files could be correctly predicted by all models.

For each file, the comments were extracted using regular expressions. After extracting the comments, the langdetect toolkit~\citep{ercdidip2022} was used to verify that the comments were written in the appropriate language. From the remaining files, comments with less than $10$ tokens in all tokenizers were filtered out. \emsemaksym{To filter out the comments which were references to licenses, TODOs, contained only the authors' personal/contact information, or were auto-generated, we randomly sampled $550$ files from the obtained set, and experts reviewed them manually. We rejected the files that contained more than one natural language, and also confirmed the presence of the correct target language detected by the langdetect.
Finally, from the files that remained after the manual verification, the first $500$ were taken for each language, resulting in $2,500$ generated comments for each language.}

\subsection{Comment Generation}
\label{sec:setup-generation}
To control the number of variables to account for, such as finetuning tasks and prompting template, we evaluate base models for their ability to generate comments. To do this, we use the Fill In the Middle~\citep{FIM} (FIM) input format. This is the same task that is used to train the base LLMs, giving us the best idea of their core performance. In order to ensure that the models will generate predictions in the correct language, we add the first 3 tokens of the ground truth in order to prime the model for the correct language. Not doing so results in almost exclusively English comment generation. To give an overview of the input format, we give an example in Figure~\ref{fig:input_format}.

\section{Qualitative investigation}
\subsection{Approach}
\paragraph{Open Coding}
To label the generations, an open coding methodology was applied. An initial set of errors based on previous work~\citep{code4me, 8544345, mariana2014multidimensional} was selected and iteratively improved in \emsemaksym{5 in-person rounds of discussion between $6$ authors, until the saturation was reached.} For each language, there was at least one author who had a native speaker level of understanding of the language to ensure that linguistic nuances, grammar correctness, and semantic accuracy were accurately evaluated.

An initial subset of $200$ files per language was selected for analysis, similar to the final dataset. The generated comments were manually compared to the originals to classify any possible errors according to the initial hypothetical error taxonomy. The results of this iteration of labeling were discussed in meetings among $6$ authors, resulting in adjustments to the error taxonomy.
After each iteration, all of the authors discussed their findings regarding the error taxonomy and adapted the taxonomy to reflect any changes that needed to be made according to the newly identified categories. In the process of adapting the taxonomy, inclusion and exclusion criteria were defined for each category to clarify when an error should be labeled within that category. \emsemaksym{To maintain the quality of the taxonomy, we ensured that the criteria were mutually exclusive and collectively exhaustive where possible.}
This evaluation process was repeated with new data until no further adjustments were suggested to the error taxonomy, resulting in five iterations until the error taxonomy no longer changed. \emsemaksym{Therefore, the saturation was reached when existing inclusion/exclusion criteria successfully accounted for all observed samples in all languages. Conflicts were resolved during the discussions by adjusting the inclusion/exclusion criteria to be language-agnostic. The criteria were generalized to focus on the underlying logical failure rather than the linguistic layer in a certain language.} The process of generating a final taxonomy and labeling all files costs $500$ person-hours.

The resulting error taxonomy was then used for error labeling in our final dataset. This dataset contains $500$ files per language for a total of $2,500$ files, with an inference from each model resulting in $5$ inferences per file. This totals $12,500$ labeled comments.
 
\paragraph{Expert Accuracy}
In addition to recording the errors made by a model, we also use an evaluation of a native speaker with programming experience to determine whether a prediction is correct, partially correct, or completely incorrect. In practice, model suggestions for code completion may be accepted, but will be edited afterward to better fit the problem at hand~\citep{10.1145/3613904.3641936, 10.1145/3597503.3608128, code4me}. Including a partially correct label gives us a more nuanced approach for situations where the models are close to a valid answer but still make minor mistakes. This accounts for developers accepting partially correct completions that they later edit themselves~\citep{code4me}.

We define the categories of correct, partially correct, and incorrect as follows. A correct prediction must include all of the information in the original comment, but may also include additional, relevant, information that was omitted. A partially correct comment is a comment where the content is largely correct but there is a small error, such as an incorrect variable name or spelling mistake. All other comments were considered incorrect.

\subsection{Results}
\paragraph{Error Taxonomy}
From the open coding investigation, we arrive at $4$ main categories of error in comment generation. We present these categories and all their subcategories in Table~\ref{tab:taxonomy}. We use this taxonomy to answer RQ1.1, What types of errors do LLMs make when generating comments in different
languages? Semantic errors, the most common category, refer to errors that relate to the meaning of the comment. This can relate to the amount of detail provided to the model, whether the model is hallucinating, and whether code is included in/after the comment. Linguistic errors refer to errors made in the language itself. This can refer to grammatical errors, but also to the synonyms the model uses and whether it responds in the correct language. Model-specific errors are errors due to the behavior of LLMs; this focuses on  memorization, copying information from the context, and knowing when to stop.

Each of the four categories can be divided into subcategories. The most common error in the Model-Specific category is the Late Termination (MS-LT) error, which occurs when the model continues to produce additional content after the relevant part of the comment has been completed. On the other hand, there is Early Termination (MS-ET), which means that the prediction stops when the comment is not complete. Other error types are Copy Context (MS-CC) when the model copies the surrounding context verbatim, and Repetition (MS-RE), when the model repeats what it has already generated, this can happen in two ways. The model can repeat a pattern, but make minor changes such as increasing a number (MS-RE1) or the model can repeat a token or set of tokens exactly the same (MS-RE2). Incoherent Generation (MS-IG) occurs when the generation consists of random words or symbols with no logic between them. Memorization (MS-ME) is assigned when the model returns personally identifiable information (MS-ME1), URL (MS-ME2), or the prediction is the same as the erroneous ground truth (MS-ME3), indicating that the comment was in the training set (e.g. the original comment contained a grammatical error and a model repeated that, or the original comment did not refer to the code correctly and the model predicted exactly the same). 

Linguistic errors are divided into three subcategories. The first is Grammar (LG-GR), where all grammatical errors are assigned. The second one, Incorrect Synonym (LG-IS), occurs when a model uses a similar word with an incorrect meaning in context. The last category is the Wrong Language (LG-WL) category. Due to the use of English loan words in many languages, the models will attempt to translate them when they should not (LG-WL1). Similarly, the model can also generate most or a significant part of a comment in a language other than the target language (LG-WL2). 

In the Semantic group, we distinguish several errors, such as the prediction with Missing Details (SE-MD) or being Too Specific (SE-TS). Many predictions also include code, either executable (SE-CS2) or commented out (SE-CS1). The subcategory for Hallucination generations is divided into three types of errors. Misplaced Facts (SE-HA1) occur when random facts that do not align with the expected content are present. Out of Context (SE-HA2), which is a hallucination not grounded in the provided context. and In Context (SE-HA3) Hallucinations, which are educated guesses based on the provided context. Finally, we have the Syntax errors (ST), which are all errors related to the syntax of the comments.
\begin{table*}[]
\centering
\begin{minipage}{13cm}
\caption{Taxonomy of common errors found during the open coding process, all languages are listed separately, and summed up in the total. Sub-categories are accumulated in the count of the parent category.}
\resizebox{\columnwidth}{!}{
\begin{tabular}{lr|ccccc}
\toprule
\textbf{Failure category with label ID} & \textbf{Total} & \textbf{Chinese} & \textbf{Dutch} & \textbf{English} & \textbf{Greek} & \textbf{Polish}\\
\midrule
\DTsetlength{0.2em}{0.7em}{0.2em}{0.4pt}{0pt}
\begin{minipage}{6cm}\dirtree{
.1 \textbf{MS Model-specific Errors}.
.2 MS-IG Incoherent Generation.
.2 MS-CC Copy context.
.2 MS-ME Memorization.
.3 MS-ME1 PII.
.3 MS-ME2 URL.
.3 MS-ME3 Training Data.
.2 MS-ET Early Termination.
.2 MS-LT Late Termination.
.2 MS-RE Repetition.
.3 MS-RE1 Pattern Repetition.
.3 MS-RE2 Verbatim Repetition.
}\end{minipage}
&
\DTsetlength{0pt}{0pt}{0pt}{0pt}{0pt}
\begin{minipage}{0.8cm}\dirtree{
.1 \rightline{\textbf{5,007}}.
.1 \rightline{259}.
.1 \rightline{991}.
.1 \rightline{422}.
.1 \rightline{236}.
.1 \rightline{114}.
.1 \rightline{72}.
.1 \rightline{164}.
.1 \rightline{2,324}.
.1 \rightline{847}.
.1 \rightline{317}.
.1 \rightline{530}.
}
\end{minipage}
&
\DTsetlength{0pt}{0pt}{0pt}{0pt}{0pt}
\begin{minipage}{1cm}\dirtree{
.1 \rightline{1,191}.
.1 \rightline{36}.
.1 \rightline{81}.
.1 \rightline{224}.
.1 \rightline{71}.
.1 \rightline{99}.
.1 \rightline{54}.
.1 \rightline{34}.
.1 \rightline{660}.
.1 \rightline{156}.
.1 \rightline{55}.
.1 \rightline{101}.
}
\end{minipage}
&
\DTsetlength{0pt}{0pt}{0pt}{0pt}{0pt}
\begin{minipage}{1cm}\dirtree{
.1 \rightline{1,072}.
.1 \rightline{42}.
.1 \rightline{281}.
.1 \rightline{40}.
.1 \rightline{36}.
.1 \rightline{1}.
.1 \rightline{3}.
.1 \rightline{48}.
.1 \rightline{409}.
.1 \rightline{238}.
.1 \rightline{106}.
.1 \rightline{132}.
}
\end{minipage}
&
\DTsetlength{0pt}{0pt}{0pt}{0pt}{0pt}
\begin{minipage}{1cm}\dirtree{
.1 \rightline{752}.
.1 \rightline{2}.
.1 \rightline{70}.
.1 \rightline{74}.
.1 \rightline{68}.
.1 \rightline{5}.
.1 \rightline{1}.
.1 \rightline{13}.
.1 \rightline{491}.
.1 \rightline{102}.
.1 \rightline{79}.
.1 \rightline{23}.
}
\end{minipage}
&
\DTsetlength{0pt}{0pt}{0pt}{0pt}{0pt}
\begin{minipage}{1cm}\dirtree{
.1 \rightline{928}.
.1 \rightline{165}.
.1 \rightline{300}.
.1 \rightline{40}.
.1 \rightline{33}.
.1 \rightline{3}.
.1 \rightline{4}.
.1 \rightline{50}.
.1 \rightline{284}.
.1 \rightline{89}.
.1 \rightline{6}.
.1 \rightline{83}.
}
\end{minipage}
&
\DTsetlength{0pt}{0pt}{0pt}{0pt}{0pt}
\begin{minipage}{1cm}\dirtree{
.1 \rightline{1,040}.
.1 \rightline{12}.
.1 \rightline{242}.
.1 \rightline{47}.
.1 \rightline{31}.
.1 \rightline{6}.
.1 \rightline{10}.
.1 \rightline{19}.
.1 \rightline{472}.
.1 \rightline{248}.
.1 \rightline{68}.
.1 \rightline{180}.
}
\end{minipage}
\\
\midrule
\DTsetlength{0.2em}{0.8em}{0.2em}{0.4pt}{0pt}
\begin{minipage}{7cm}
\dirtree{%
.1 \textbf{LG Linguistic Error}.
.2 LG-GR Grammar.
.3 LG-GR1 Plurality. 
.3 LG-GR2 Conjugation.
.3 LG-GR3 Gender.
.3 LG-GR4 Language Syntax.
.3 LG-GR5 Cohesion.
.2 LG-IS Incorrect synonym.
.2 LG-WL Wrong language.
.3 LG-WL1 Undesired translations.
.3 LG-WL2 Incorrect language.
}
\end{minipage}
&
\DTsetlength{0pt}{0pt}{0pt}{0pt}{0pt}
\begin{minipage}{0.8cm}
\dirtree{%
.1 \rightline{\textbf{1,728}}.
.1 \rightline{1,420}.
.1 \rightline{17}.
.1 \rightline{169}.
.1 \rightline{130}.
.1 \rightline{227}.
.1 \rightline{876}.
.1 \rightline{128}.
.1 \rightline{180}.
.1 \rightline{40}.
.1 \rightline{140}.
}
\end{minipage}
&
\DTsetlength{0pt}{0pt}{0pt}{0pt}{0pt}
\begin{minipage}{1cm}
\dirtree{%
.1 \rightline{17}.
.1 \rightline{0}.
.1 \rightline{0}.
.1 \rightline{0}.
.1 \rightline{0}.
.1 \rightline{0}.
.1 \rightline{0}.
.1 \rightline{0}.
.1 \rightline{17}.
.1 \rightline{1}.
.1 \rightline{16}.
}
\end{minipage}
&
\DTsetlength{0pt}{0pt}{0pt}{0pt}{0pt}
\begin{minipage}{1cm}
\dirtree{%
.1 \rightline{220}.
.1 \rightline{124}.
.1 \rightline{1}.
.1 \rightline{8}.
.1 \rightline{20}.
.1 \rightline{25}.
.1 \rightline{70}.
.1 \rightline{0}.
.1 \rightline{96}.
.1 \rightline{8}.
.1 \rightline{88}.
}
\end{minipage}&
\DTsetlength{0pt}{0pt}{0pt}{0pt}{0pt}
\begin{minipage}{1cm}
\dirtree{%
.1 \rightline{66}.
.1 \rightline{65}.
.1 \rightline{0}.
.1 \rightline{0}.
.1 \rightline{0}.
.1 \rightline{42}.
.1 \rightline{23}.
.1 \rightline{0}.
.1 \rightline{1}.
.1 \rightline{0}.
.1 \rightline{1}.
}
\end{minipage}
&
\DTsetlength{0pt}{0pt}{0pt}{0pt}{0pt}
\begin{minipage}{1cm}
\dirtree{%
.1 \rightline{979}.
.1 \rightline{870}.
.1 \rightline{14}.
.1 \rightline{59}.
.1 \rightline{82}.
.1 \rightline{38}.
.1 \rightline{677}.
.1 \rightline{95}.
.1 \rightline{14}.
.1 \rightline{12}.
.1 \rightline{2}.
}
\end{minipage}
&
\DTsetlength{0pt}{0pt}{0pt}{0pt}{0pt}
\begin{minipage}{1cm}
\dirtree{%
.1 \rightline{423}.
.1 \rightline{350}.
.1 \rightline{1}.
.1 \rightline{102}.
.1 \rightline{23}.
.1 \rightline{121}.
.1 \rightline{103}.
.1 \rightline{32}.
.1 \rightline{41}.
.1 \rightline{18}.
.1 \rightline{23}.
}
\end{minipage}
\\
\midrule
\DTsetlength{0.2em}{0.7em}{0.2em}{0.4pt}{0pt}
\begin{minipage}{5.5cm}
\dirtree{%
.1 \textbf{SE Semantic error}.
.2 SE-MD Missing Details.
.2 SE-TS Too Specific.
.2 SE-HA Hallucination.
.3 SE-HA1 Misplaced Facts.
.3 SE-HA2 Out of Context.
.3 SE-HA3 In context.
.2 SE-CS  Code Inclusion.
.3 SE-CS1 Commented code.
.3 SE-CS2 Runnable code.
.2 SE-OI Omitted Identifier.
}
\end{minipage}
&
\DTsetlength{0pt}{0pt}{0pt}{0pt}{0pt}
\begin{minipage}{0.8cm}
\dirtree{%
.1 \rightline{\textbf{8,333}}.
.1 \rightline{837}.
.1 \rightline{168}.
.1 \rightline{4,239}.
.1 \rightline{325}.
.1 \rightline{550}.
.1 \rightline{3,364}.
.1 \rightline{3,009}.
.1 \rightline{171}.
.1 \rightline{2,838}.
.1 \rightline{80}.
}
\end{minipage}
&
\DTsetlength{0pt}{0pt}{0pt}{0pt}{0pt}
\begin{minipage}{1cm}
\dirtree{%
.1 \rightline{2,915}.
.1 \rightline{413}.
.1 \rightline{91}.
.1 \rightline{1,639}.
.1 \rightline{79}.
.1 \rightline{385}.
.1 \rightline{1,175}.
.1 \rightline{718}.
.1 \rightline{6}.
.1 \rightline{712}.
.1 \rightline{54}.
}
\end{minipage}
&
\DTsetlength{0pt}{5pt}{0pt}{0pt}{0pt}
\begin{minipage}{1cm}
\dirtree{%
.1 \rightline{1,381}.
.1 \rightline{106}.
.1 \rightline{21}.
.1 \rightline{664}.
.1 \rightline{102}.
.1 \rightline{78}.
.1 \rightline{484}.
.1 \rightline{578}.
.1 \rightline{66}.
.1 \rightline{512}.
.1 \rightline{12}.
}
\end{minipage}
&
\DTsetlength{0pt}{0pt}{0pt}{0pt}{0pt}
\begin{minipage}{1cm}
\dirtree{%
.1 \rightline{786}.
.1 \rightline{19}.
.1 \rightline{1}.
.1 \rightline{406}.
.1 \rightline{42}.
.1 \rightline{12}.
.1 \rightline{352}.
.1 \rightline{355}.
.1 \rightline{12}.
.1 \rightline{343}.
.1 \rightline{5}.
}
\end{minipage}
&
\DTsetlength{0pt}{0pt}{0pt}{0pt}{0pt}
\begin{minipage}{1cm}
\dirtree{%
.1 \rightline{1,726}.
.1 \rightline{95}.
.1 \rightline{12}.
.1 \rightline{913}.
.1 \rightline{7}.
.1 \rightline{57}.
.1 \rightline{849}.
.1 \rightline{706}.
.1 \rightline{14}.
.1 \rightline{792}.
.1 \rightline{0}.
}
\end{minipage}
&
\DTsetlength{0pt}{0pt}{0pt}{0pt}{0pt}
\begin{minipage}{1cm}
\dirtree{%
.1 \rightline{1,523}.
.1 \rightline{202}.
.1 \rightline{42}.
.1 \rightline{630}.
.1 \rightline{91}.
.1 \rightline{16}.
.1 \rightline{523}.
.1 \rightline{640}.
.1 \rightline{70}.
.1 \rightline{570}.
.1 \rightline{9}.
}
\end{minipage}
\\
\midrule
\DTsetlength{0.2em}{0.7em}{0.2em}{0.4pt}{0pt}
\begin{minipage}{7cm}
\dirtree{%
.1 \textbf{ST Syntax}.
.2 ST-IF Incorrect comment format.
}
\end{minipage}
&
\DTsetlength{0pt}{0pt}{0pt}{0pt}{0pt}
\begin{minipage}{1cm}
\dirtree{%
.1 \rightline{\textbf{84}}.
.1 \rightline{84}.
}
\end{minipage}
&
\DTsetlength{0pt}{0pt}{0pt}{0pt}{0pt}
\begin{minipage}{1cm}
\dirtree{%
.1 \rightline{31}.
.1 \rightline{31}.
}
\end{minipage}
&
\DTsetlength{0pt}{0pt}{0pt}{0pt}{0pt}
\begin{minipage}{1cm}
\dirtree{%
.1 \rightline{2}.
.1 \rightline{2}.
}
\end{minipage}&
\DTsetlength{0pt}{0pt}{0pt}{0pt}{0pt}
\begin{minipage}{1cm}
\dirtree{%
.1 \rightline{21}.
.1 \rightline{21}.
}
\end{minipage}
&
\DTsetlength{0pt}{0pt}{0pt}{0pt}{0pt}
\begin{minipage}{1cm}
\dirtree{%
.1 \rightline{5}.
.1 \rightline{5}.
}
\end{minipage}
&
\DTsetlength{0pt}{0pt}{0pt}{0pt}{0pt}
\begin{minipage}{1cm}
\dirtree{%
.1 \rightline{25}.
.1 \rightline{25}.
}
\end{minipage}
\\

\bottomrule
\end{tabular}
}
\label{tab:taxonomy}
\end{minipage}
\end{table*}

This taxonomy shows that the main issue with comments generated by LLMs is Semantic Errors, often in the form of Hallucinations (SE-HA). These errors account for the largest increase in errors for non-English languages compared to English. Furthermore, we see that the level of detail (SE-MD and SE-TS) is a cause of errors mainly in non-English generations. Looking at the Model Specific errors (MS) we see that for errors related to Memorization (MS-ME), the models are most likely to make them in English compared to other Western languages. Interestingly, Chinese has the highest rate of memorization for all categories. For language-specific errors (LG), we see that Chinese does not have grammar errors, as discussed in Section~\ref{non-english}. Furthermore, we notice that Dutch has the highest likelihood of returning predictions in the wrong language. We believe this is due to the similarity of Dutch to English and the high amount of overlap between the two dictionaries. We also observe a high rate of incorrect synonyms used in Polish compared to all other languages and a high rate of language syntax errors for English compared to all other language errors made by the model. Finally, in the Syntax errors (ST) we see that the models' ability to adhere to comment syntax is dependent on the language it is using.

We further compare languages on a per-language basis and see that Greek has the highest number of errors for the language grammar, as well as the use of incorrect synonyms. This amounts to an increase of $15.1\times$ the error rate compared to English. This also appears in the MS category, where Greek has the highest tendency to generate incoherent output.
However, Greek is the language that is less likely to generate repetitions of all languages analyzed. We also see that Dutch, Greek, and Polish are more likely to copy the surrounding context into the comment, while Chinese do it at a rate similar to English.

\begin{table*}
\centering
\caption[Expert accuracy of all models and languages]{Expert accuracy for all models and languages ($\checkmark$ correct, $\thicksim$ partial, $\mytimes$ incorrect)}
\label{tab:expert_accuracy}
\resizebox{\columnwidth}{!}{
\begin{tabular}{c|rrr|rrr|rrr|rrr|rrr}
\multicolumn{1}{l}{} & \multicolumn{3}{c}{CodeGemma} & \multicolumn{3}{c}{CodeLlama} & \multicolumn{3}{c}{CodeQwen 1.5} & \multicolumn{3}{c}{GraniteCode} & \multicolumn{3}{c}{StarCoder2} \\
\multicolumn{1}{l|}{} & \multicolumn{1}{c}{$\checkmark$} & \multicolumn{1}{c}{$\thicksim$} & \multicolumn{1}{c|}{$\mytimes$} & \multicolumn{1}{c}{$\checkmark$} & \multicolumn{1}{c}{$\thicksim$} & \multicolumn{1}{c|}{$\mytimes$} & \multicolumn{1}{c}{$\checkmark$} & \multicolumn{1}{c}{$\thicksim$} & \multicolumn{1}{c|}{$\mytimes$} & \multicolumn{1}{c}{$\checkmark$} & \multicolumn{1}{c}{$\thicksim$} & \multicolumn{1}{c|}{$\mytimes$} & \multicolumn{1}{c}{$\checkmark$} & \multicolumn{1}{c}{$\thicksim$} & \multicolumn{1}{c}{$\mytimes$} \\
\hline
Chinese & 55 & 221 & 224 & 63 & 201 & 236 & 84 & 200 & 216 & 67 & 175 & 258 & 20 & 218 & 262 \\
Dutch & 298 & 98 & 104 & 246 & 95 & 159 & 158 & 109 & 233 & 253 & 80 & 167 & 215 & 107 & 178 \\
English & 406 & 76 & 18 & 266 & 162 & 72 & 268 & 183 & 49 & 339 & 123 & 38 & 320 & 145 & 35 \\
Greek & 259 & 117 & 124 & 118 & 119 & 263 & 78 & 165 & 257 & 132 & 130 & 238 & 118 & 131 & 251 \\
Polish & 271 & 146 & 83 & 176 & 164 & 160 & 159 & 157 & 184 & 205 & 151 & 144 & 167 & 154 & 179 \\
\end{tabular}
}
\end{table*}

\subsection{Expert Accuracy}
To understand how well the chosen models perform in generating code comments in languages they were not intended to be used for (RQ1.2), the experts labeled each generation as either \texttt{Correct}, \texttt{Partially Correct}, or \texttt{Incorrect}. We give an overview of the scores in Table~\ref{tab:expert_accuracy}. Here we see that while models claim to be only intended to be used for English, and sometimes Chinese, we see that they generate at least partially accurate completions in more than half of the cases for Dutch, Greek, and Polish. Interestingly, we see that although Chinese is the second most common language present in training corpora, it is the worst-performing language overall. On a model level, we see that most models follow the same trend, they perform best in English, then Dutch or Polish, followed by Greek, and finally Chinese. However, for CodeGemma, we see that the performance of Greek is doubled in comparison to other models.

\section{Off-the-shelf metrics}
\label{sec:off-the-shelf}
\subsection{Approach}
\label{sec:off-the-shelf_approach}
For the numerical analysis of the comments and the validation of the metrics, we treat the problem as a text generation problem. We have a prediction given by the model and a reference comment that we compare it to. For this evaluation, we use a combination of word level and neural metrics. To validate the ability of metrics to score a prediction, we focus on the requirement of a metric to significantly reduce the score of a generation due to a perturbation~\citep{sai-etal-2021-perturbation}. We push these perturbations to the extreme. We generate two types of random noise. First we generate ``uniform noise'' which is any token from the tokenizer of the models, equal in length to the target comment in tokens, and ``targeted noise'' which are random tokens sampled from the surrounding context.

\paragraph{Word Level Metrics}
For word-level metrics, we use the standard metrics used in code-related studies: BLEU~\citep{papineni2002bleu}, ROUGE~\citep{lin2004rouge}, and METEOR~\citep{banerjee2005meteor}. BLEU measures precision by comparing n-gram matches between the generated and reference summaries while penalizing overgeneration with a brevity penalty. ROUGE focuses on recall by measuring the overlap of n-grams between the reference and the generated texts. ROUGE-1 considers unigrams, ROUGE-2 considers bi-grams, and ROUGE-L emphasizes the longest common subsequence to assess fluency and logical consistency. For this investigation, we focus on ROUGE-L.  METEOR combines precision, recall, and synonym-based matching to provide the score. 

\paragraph{Embedding Based Metrics}
These metrics calculate the cosine similarity between embeddings. BERTScore~\citep{zhang2019bertscore} uses a BERT~\citep{kenton2019bert} model to generate the embeddings. The BERT configuration varies according to the target language. For English tasks, a 24-layer RoBERTa large model is used, while for Chinese tasks, a 12-layer BERT Chinese model is utilized. For other languages, it relies on the 12-layer cased multilingual BERT (BERT$_{multi}$) model. BERT$_{multi}$ was trained in 104 languages using the Wikipedia corpora, including Greek, Polish, and Dutch. However, this model has not been trained on code. CodeBERTScore~\citep{zhou2023codebertscore} is similar to BERTScore, however, it was specifically developed for code evaluation, it differs from BERTScore, by using the CodeBERT model~\citep{feng2020codebert} and ignoring some keywords and syntax tokens common throughout the code.

\paragraph{Model Based Metrics} BARTScore~\citep{yuan2021bartscore}
employs BART~\citep{lewis2019bart}, a transformer model that evaluates the quality of the generated text through the likelihood scores assigned by the model. The BART models are trained by both corrupting the input, and introducing arbitrary noise into the input text, after which it reconstructs the original input.
Although specific details regarding training data are not explicitly provided, some indirect information suggests that it includes datasets such as books and Wikipedia entries—similar to those used in RoBERTa's pre-training process~\citep{liu2019roberta}. Similar to BERTScore, code is not specifically mentioned in the training procedure.

\begin{figure*}
    \centering
    \includegraphics[width=1\linewidth]{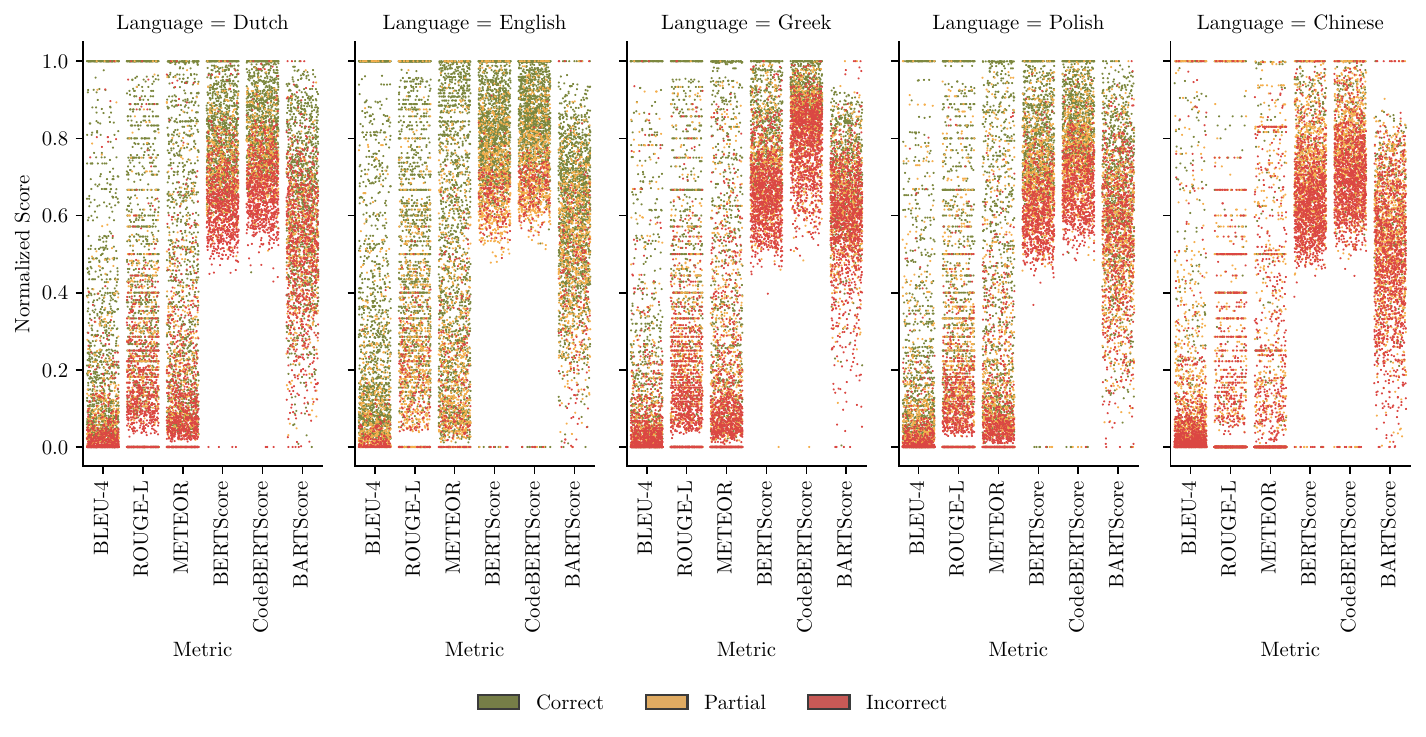}
    \caption{Strip plot, showing the scores assigned to comment generations by different metrics.}
    \label{fig:neural_eval}
\end{figure*}

\begin{figure*}
    \centering
    \includegraphics[width=1\linewidth]{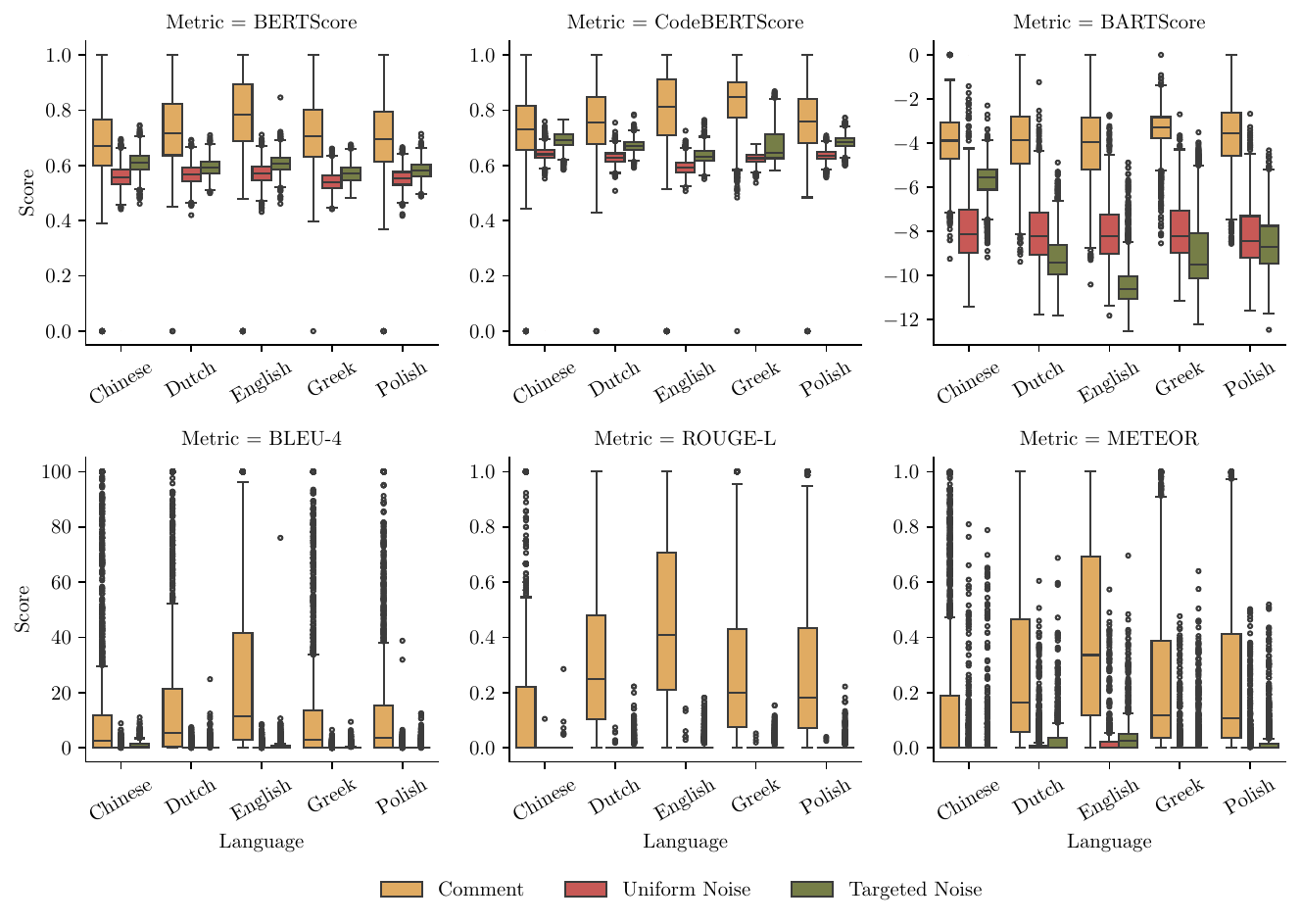}
    \caption{Metric scores comparing LLM generated comments, to random samples of tokens using two separate distributions.}
    \label{fig:metrics}
\end{figure*}

\subsection{Results}
\paragraph{Scoring Collections}
We answer RQ2.1 by analyzing the scores that the selected metrics assign to both the predictions and the random noise that we generated. An overview is given in Figure~\ref{fig:metrics}. Here, we look at two factors identified as belonging to a good metric, alignment with expert evaluations, and the ability to differentiate between noise and real prediction.

Looking at only the English results, we see that the scores assigned by the embedding based metrics align closely with the expert evaluation (around $0.8$). However, when looking at other languages, the embedding-based metrics score the predictions slightly lower, even when the expert evaluation scores the generations much lower (especially for Chinese). However, word-based metrics on the other hand score predictions lower in general but do match the differences in performance observed in the expert accuracy.

\paragraph{Differentiating Correct From Incorrect}
To answer RQ2.2, Do common evaluation metrics effectively differentiate between correct and incorrect predictions? We plot the scores of selected metrics in Figure~\ref{fig:neural_eval}. Here we show a strip plot of the score for each generated comment and color it by the expert evaluation \texttt{Correct} (green), \texttt{Partially Correct} (orange), and \texttt{Incorrect} (red). Each point in a stripe represents a single generation. To demonstrate the spreads of different scores, we normalize all scores between $0$ and $1$. This shows us if there is a clear separation between the three colors at a glance.
We see a difference in the distribution between neural models and classical models. Classical models have the majority of their data points scored low and have a tail of scores that score high. On the other hand, neural metrics, especially the embedding-based metrics, have their scores concentrated towards the top, with few points in the lower half of the plot. Looking at all the strip plots, we see no clear separation between correct, partially correct, or incorrect predictions. Although there is a slight trend to score the correct predictions higher, a large overlap between the three colors persists, showing that the metrics do not separate them accurately.

\paragraph{Robustness to Noise}
Finally, we look at the generated noise to answer RQ2.3, we see that for all neural metrics (in Figure~\ref{fig:metrics}, there is a significant overlap between the scores assigned to noise and the scores assigned to real generations. The BARTScores have a lower overlap compared to the embedding based metrics; however, interestingly, the targeted noise performed worse than the uniform noise for all languages except Chinese. Finally, we see a general trend of CodeBERTScore scoring predictions higher than BERTScore, hinting that the training procedure is a more important factor for a higher BERTScore than the generations being studied. For classic metrics, there is a far lower overlap between noise and real comments, and most noise is scored close to $0$ with the METEOR score showing the largest number of outliers towards high scores.

\section{Extended neural metrics}
\emsejon{While embedding-based and model-based metrics are widely adopted as off-the-shelf evaluation tools, they are typically deployed with a default model and operate under limited contextual assumptions. In particular, the context available during scoring often does not capture all information required to assess semantic correctness in code comments. To address this limitation, we extend existing metrics by systematically varying the amount of contextual information provided during evaluation, while preserving their underlying scoring mechanisms. This is motivated by the observation that semantically relevant information for a code comment is often located in the surrounding code, and cannot be inferred from the comment alone.}

\emsejon{In addition, as observed in Section~\ref{sec:off-the-shelf}, neural metrics can produce score distributions that do not align with human judgments. For example, non-English generations may receive higher scores than English ones despite being rated lower by human experts. This highlights that absolute metric scores are not reliable indicators of generation quality. We therefore introduce a comparison framework based on agreement with human annotations, rather than raw metric scores.}

\emsejon{We evaluate the extended metrics in terms of their alignment with human judgment and their robustness to perturbations, as introduced in Section~\ref{sec:off-the-shelf_approach}.}

\subsection{Approach}
\emsejon{We evaluate extended neural metrics using a three-stage pipeline consisting of (i) input generation with varying levels of context, (ii) metric computation, and (iii) comparison with human annotations through a classification-based framework.}

\paragraph{Input generation}
\emsejon{When working with neural models that have an encoder or encoder/decoder based architectures, they generate outputs based on the context that surrounds the target tokens. This means that changing the context around a generation will influence the embeddings/loss of a model. }

\emsejon{For code comment generation, information relevant to semantic correctness is often located in the surrounding code, either preceding or most commonly following the comment. However, existing implementations of neural evaluation metrics do not support incorporating such context. For example, BERTScore and BARTScore operate solely on the generated text~\citep{zhang2019bertscore, yuan2021bartscore}, while CodeBERTScore allows only limited prefix context~\citep{zhou2023codebertscore}, which does not capture code that follows the comment.}

\emsejon{To study the effect of contextual information, we extend BERTScore and BARTScore with three context settings:}
\begin{itemize}
    \item \emsejon{\textbf{No context:} Only the generated comment tokens and their corresponding ground truth are provided, without additional context. This results in partial or incomplete sentences at the start of the comment and no information about the surrounding code.}

    \item \emsejon{\textbf{Minimal context:} Only the generated comment is provided, along with a short prefix indicating the target language (as described in Section~\ref{sec:setup-generation}). This ensures well-formed input while removing most
    contextual information.}

    \item \emsejon{\textbf{Full context:} The entire file is provided as input, with the generated comment inserted at the correct location. This setting provides complete contextual information and is expected to support detection of semantic errors.}
\end{itemize}

\begin{figure}
    \centering
    \includegraphics[width=1.1\linewidth]{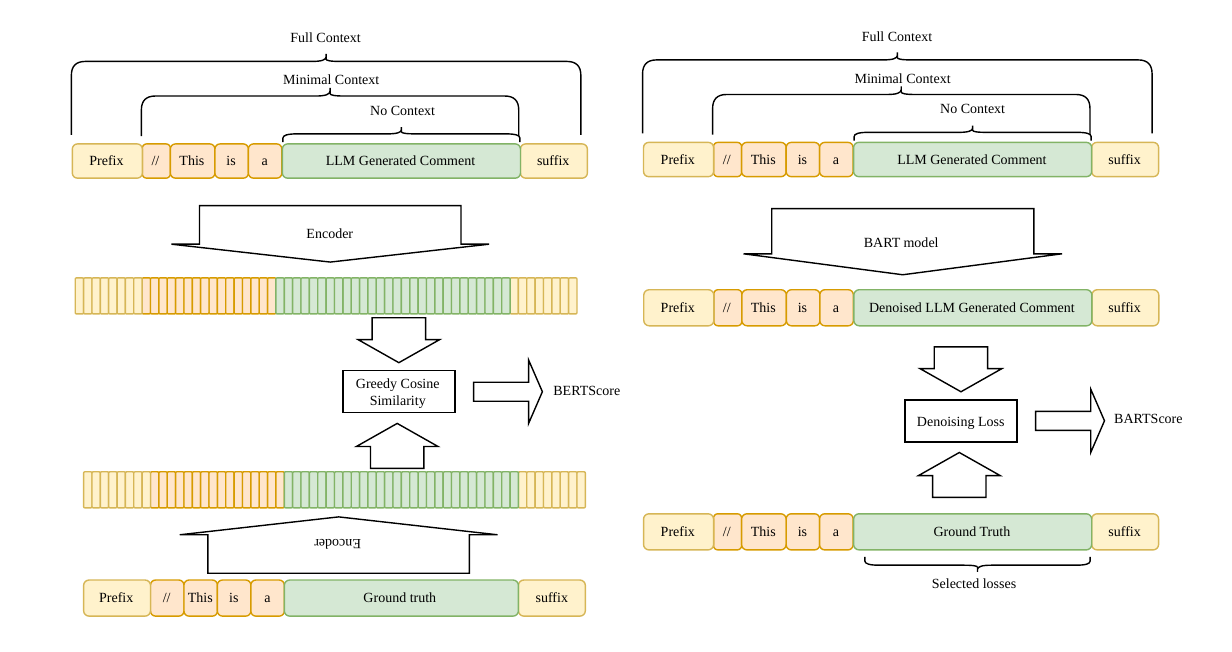}
    \caption{A visual demonstration of how different amounts of context will affect the neural metrics.}
    \label{fig:extended_metric}
\end{figure}

\paragraph{Metric Computation}
\emsejon{For the remainder of this investigation, we refer to all encoder-based metrics under the unified term \textit{BERTScore}. BERTScore and CodeBERTScore differ primarily in three aspects. First, CodeBERTScore allows for the inclusion of a short natural language prefix as additional input. Second, CodeBERTScore was designed for code inputs and therefore excludes syntax tokens from the similarity computation. Third, the underlying encoder model differ depending on the implementation, CodeBERTScore focuses on code models, while BERTScore focuses on natural language models. In our setting, including a small prefix of text before the similarity computation is equivalent to the minimal context setting, our input is predominantly natural languages due to it being code comments, reducing the need to filter out syntax tokens, and finally, we run our experiments for multiple code and natural language encoder models, making the difference between CodeBERTScore and BERTScore superfluous for this setting.}

\emsejon{When computing BARTScores and BERTScores, we use all tokens relevant to the given context setting as input to the models, however, once the loss/embedding has been calculated, we exclude all context tokens and use only the tokens belonging to the generated comment and the ground truth. This ensures that scores are not influenced by the quality of the surrounding code, while still allowing contextual information to affect the underlying representations through the model input. This process is illustrated in Figure~\ref{fig:extended_metric}.}

\paragraph{Comparison}
\label{sec:Extended_comparison}
\emsejon{The large number of models evaluated in this study requires a principled approach to comparing metric performance. As observed in Section~\ref{sec:off-the-shelf}, neural metrics produce score distributions that do not align with human judgments. For example, certain non-English generations may receive higher scores than English ones despite being rated lower by human experts, or randomly generated noise may score higher than a real prediction. This indicates that we need a method to calculate alignment with human experts, without relying on the absolute score of the metrics.}

\emsejon{To enable this comparison, we instead focus on the relationship between metric scores and human annotations. We leverage the expert labels obtained in the qualitative study (\texttt{correct}, \texttt{partially correct}, \texttt{incorrect}) and associate them with the corresponding metric scores. For each metric, we fit a kernel density estimator (KDE) to model the score distribution for each class. We then compute the intersection points between these distributions to derive threshold values that partition the score space into regions corresponding to each label. Our choice for a KDE classifier stems from the fact that we are working in a single dimension, and we need to account for the class imbalances in our set of labels.}

\begin{figure}
    \centering
    \includegraphics[width=0.9\linewidth]{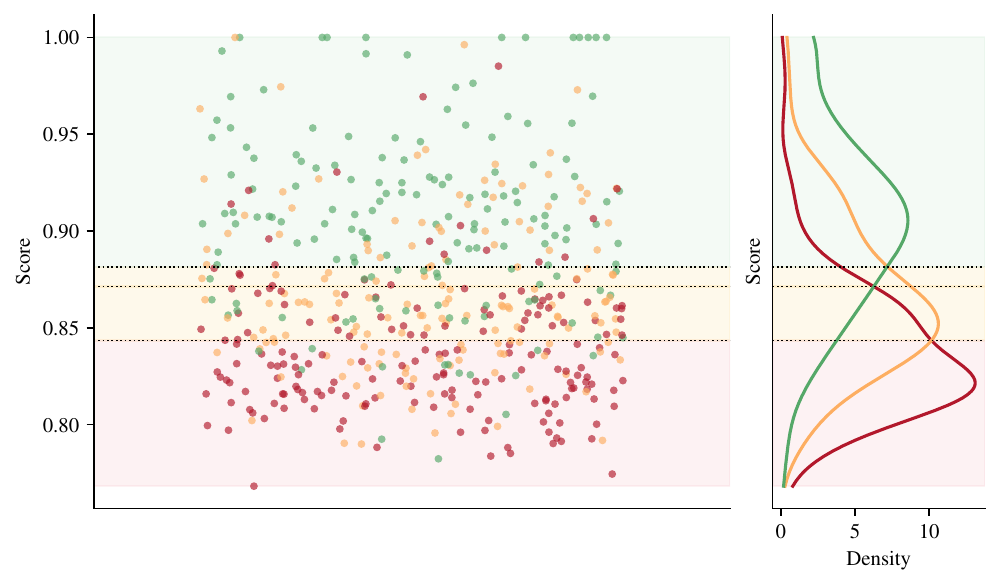}
    \caption{Example of the classifier used for label assignment, the scatter plot shows each generated comment, colored in red for an incorrect generation, yellow for a partially correct generation and green for a correct generation. The KDE-densities are plotted on the right side with the corresponding colors. All intersections are marked with a horizontal line and the background is shaded by the classifier output. The classifier assigns the label of the background color as the new label of each sample, which are then used for calculating the alignment.}
    \label{fig:metrics-classifier}
\end{figure}

\emsejon{Using these thresholds, we reclassify all samples based on their metric scores, resulting in a new set of predicted labels. This classification procedure is illustrated in Figure~\ref{fig:metrics-classifier}.}

\emsejon{Given the two sets of labels, namely human annotations and metric-based predictions, we measure agreement between them to assess metric reliability. As the labels (\texttt{correct}, \texttt{partially correct}, \texttt{incorrect}) are ordinal, misclassifications are not equally severe. For example, predicting \texttt{partially correct} instead of \texttt{correct} is less severe than predicting \texttt{incorrect}. To account for this, we use quadratically weighted Cohen’s kappa as our primary agreement measure.}

\emsejon{For noise detection, we apply the same framework using real predictions and randomly generated strings. In this setting, the task reduces to binary classification (\texttt{noise} vs.\ \texttt{not noise}). We again fit KDE distributions, derive thresholds, and evaluate agreement with ground truth using Cohen’s kappa.}

\paragraph{Model Selection}
\emsejon{When selecting models to use for the neural metrics, we prioritize a varied selection of model architectures and sizes. We also focus on having at least one member for each target language, general multilingual training, and specifically trained on code. Within these groups we aim to have different architectures such as BERT~\citep{kenton2019bert}, RoBERTA~\citep{liu2019roberta}, DeBERTA~\citep{he2021debertav3}, ELECTRA~\citep{Clark2020ELECTRA:}, XLM~\citep{lample2019cross}, for the embedding based metrics, and BART~\citep{lewis2019bart} or T5~\citep{raffel2020exploring} architectures for the model based metrics where available. However, especially for language specific models, there are very few available. Using these selection criteria, we run experiments using the models listed in Table~\ref{tab:bertscore_models} and Table~\ref{tab:BART_embedding_models}.
}

\begin{table}[h!]
    \centering
    \renewcommand{\arraystretch}{1.2} 
    \resizebox{\textwidth}{!}{
    \begin{tabular}{l|c|c|l|c}
        \hline
        \textbf{Model Identifier} & \textbf{Family} & \textbf{Size} & \textbf{Training Languages} & \textbf{Citation} \\
        \hline
        \multicolumn{5}{c}{\textit{Code-Focused Encoder Models}} \\
        \hline
        answerdotai/ModernBERT-base & ModernBERT & ~125M & English + Code & \citep{modernbert} \\
        answerdotai/ModernBERT-large & ModernBERT & ~350M & English + Code & \citep{modernbert} \\
        bigcode/starencoder & StarEncoder & 125M & Code (multi-PL) & \citep{devlin2019bertpretrainingdeepbidirectional} \\
        microsoft/codebert-base & CodeBERT & 125M & Multi (PLs + En) & \citep{feng2020codebertpretrainedmodelprogramming} \\
        huggingface/CodeBERTa-small-v1 & CodeBERTa & ~125M & Multi (PLs) & \citep{husain_codesearchnet_2019} \\
        \hline
        \multicolumn{5}{c}{\textit{Multilingual Encoder Models}} \\
        \hline
        google-bert/bert-base-multilingual-cased & mBERT & 110M & 104 languages & \citep{DBLP:journals/corr/abs-1810-04805} \\
        distilbert/distilbert-base-multilingual-cased & Distil mBERT & 134M & Multi & \citep{Sanh2019DistilBERTAD} \\
        FacebookAI/xlm-mlm-100-1280 & XLM & ~570M & 100 languages & \citep{lample2019cross} \\
        FacebookAI/xlm-mlm-17-1280 & XLM & ~570M & 17 languages & \citep{lample2019cross} \\
        FacebookAI/xlm-roberta-base & XLM-R & 270M & 100 languages & \citep{lample2019cross} \\
        FacebookAI/xlm-roberta-large & XLM-R & 550M & 100 languages & \citep{lample2019cross} \\
        microsoft/mdeberta-v3-base & mDeBERTa & ~278M & Multi & \citep{he2021debertav3} \\
        AILabTUL/mELECTRA & mELECTRA & 14,3M & Multi & \citep{polacek-2025-study} \\
        
        \hline
        \multicolumn{5}{c}{\textit{English Encoder Models}} \\
        \hline
        FacebookAI/xlm-mlm-en-2048 & English XLM & 665M & English & \citep{lample2019cross} \\
        google-bert/bert-base-uncased & BERT & 110M & English & \citep{devlin2019bertpretrainingdeepbidirectional} \\
        google-bert/bert-large-uncased & BERT & 340M & English & \citep{devlin2019bertpretrainingdeepbidirectional} \\
        albert/albert-base-v2 & ALBERT & 12M & English & \citep{Lan2020ALBERT} \\
        albert/albert-large-v2 & ALBERT & 18M & English & \citep{Lan2020ALBERT} \\
        microsoft/deberta-v3-base & DeBERTa-v3 & 98M & English & \citep{he2021debertav3} \\
        microsoft/deberta-v3-large & DeBERTa-v3 & 131M & English & \citep{he2021debertav3} \\
        answerdotai/ModernBERT-base & ModernBERT & ~125M & English + Code & \citep{modernbert} \\
        answerdotai/ModernBERT-large & ModernBERT & ~350M & English + Code & \citep{modernbert} \\
        \hline
        \multicolumn{5}{c}{\textit{Chinese Encoder Models}} \\
        \hline
        hfl/chinese-macbert-base & MacBERT & ~102M & Chinese & \citep{cui-etal-2020-revisiting} \\
        hfl/chinese-macbert-large & MacBERT & ~324M & Chinese & \citep{cui-etal-2020-revisiting} \\
        hfl/chinese-roberta-wwm-ext & RoBERTa & ~102M & Chinese & \citep{cui-etal-2020-revisiting} \\
        hfl/chinese-bert-wwm-ext & BERT & ~102M & Chinese & \citep{cui-etal-2020-revisiting} \\
        \hline
        \multicolumn{5}{c}{\textit{Dutch Encoder Models}} \\
        \hline
        GroNLP/bert-base-dutch-cased & BERTje & ~100M & Dutch & \citep{de2019bertje} \\
        pdelobelle/robbert-v2-dutch-base & RobBERT & ~100M & Dutch & \citep{delobelle2020robbert} \\
        DTAI-KULeuven/robbert-2023-dutch-large & RobBERT & ~400M & Dutch & \citep{delobelle2023robbert2023conversion} \\
        DTAI-KULeuven/robbert-2023-dutch-base & RobBERT & ~100M & Dutch & \citep{delobelle2023robbert2023conversion} \\
        DTAI-KULeuven/robbertje-1-gb-merged & RobBERTje & ~74M & Dutch & \citep{delobelle2023robbert2023conversion} \\
        DTAI-KULeuven/robbertje-1-gb-bort & RobBERTje & ~46M & Dutch & \citep{delobelle2023robbert2023conversion} \\
        \hline
        \multicolumn{5}{c}{\textit{Greek Encoder Models}} \\
        \hline
        nlpaueb/bert-base-greek-uncased-v1 & GreekBERT & ~110M & Greek & \citep{greek-bert} \\
        novelcore/gem-roberta & RoBERTa & ~125M & Greek & \citep{novelcore2024gemroberta} \\
        novelcore/gem-modernbert & ModernBERT & ~139M & Greek & \citep{novelcore2024gemroberta} \\
        novelcore/gem-electra & ELECTRA & ~100M & Greek & \citep{novelcore2024gemroberta} \\
        \hline
        \multicolumn{5}{c}{\textit{Polish Encoder Models}} \\
        \hline
        dkleczek/bert-base-polish-cased-v1 & PolBERT & ~110M & Polish & \citep{kleczek2020bertpolish} \\
        allegro/herbert-base-cased & HerBERT & ~100M & Polish & \citep{mroczkowski-etal-2021-herbert} \\
        allegro/herbert-base-large & HerBERT & ~400M & Polish & \citep{mroczkowski-etal-2021-herbert} \\
        sdadas/polish-roberta-base-v2 & RoBERTa & ~100M & Polish & \citep{dadas2020pre} \\
        sdadas/polish-roberta-large-v2 & RoBERTa & ~400M & Polish & \citep{dadas2020pre} \\
        \hline
    \end{tabular}%
    }
    \caption{A list of all embedding models used for the BERTScore evaluation.}
    \label{tab:bertscore_models}
\end{table}

\begin{table}[]
    \centering
    \renewcommand{\arraystretch}{1.2} 
    \resizebox{\textwidth}{!}{
    \begin{tabular}{l|c|c|l|c}
        \hline
        \textbf{Model Identifier} & \textbf{Family} & \textbf{Size} & \textbf{Training Languages} & \textbf{Citation} \\
        \hline
        \multicolumn{5}{c}{\textit{Multi-lingual Models}} \\
        \hline
        facebook/mbart-large-cc25 & mBART & 610M & 25 languages & \citep{liu2020multilingualdenoisingpretrainingneural} \\
        google/mt5-small & mT5 & 300M & 101 languages & \citep{xue-etal-2021-mt5} \\
        google/mt5-base & mT5 & 580M & 101 languages & \citep{xue-etal-2021-mt5} \\
        google/mt5-large & mT5 & 1.2B & 101 languages & \citep{xue-etal-2021-mt5} \\
        google/mt5-xl & mT5 & 3.7B & 101 languages & \citep{xue-etal-2021-mt5} \\
        google/byt5-small & ByT5 & 300M & Multilingual & \citep{xue2022byt5tokenfreefuturepretrained} \\
        google/byt5-base & ByT5 & 580M & Multilingual & \citep{xue2022byt5tokenfreefuturepretrained} \\
        google/byt5-large & ByT5 & 1.2B & Multilingual & \citep{xue2022byt5tokenfreefuturepretrained} \\
        google/byt5-xl & ByT5 & 3.7B & Multilingual & \citep{xue2022byt5tokenfreefuturepretrained} \\
        google/umt5-small & uMT5 & 300M & Multilingual & \citep{xue-etal-2021-mt5} \\
        google/umt5-base & uMT5 & 580M & Multilingual & \citep{xue-etal-2021-mt5} \\
        google/umt5-xl & uMT5 & 3.7B & Multilingual & \citep{xue-etal-2021-mt5} \\
        \hline
        \multicolumn{5}{c}{\textit{Multi-lingual Code Models}} \\
        \hline
        Salesforce/codet5-small & CodeT5 & 60M & Multi (7 PLs + En) & \citep{wang2021codet5} \\
        Salesforce/codet5-base & CodeT5 & 220M & Multi (7 PLs + En) & \citep{wang2021codet5} \\
        Salesforce/codet5-large & CodeT5 & 770M & Multi (7 PLs + En) & \citep{wang2021codet5} \\
        Salesforce/codet5p-220m & CodeT5+ & 220M & Multi (9 PLs + En) & \citep{wang2023codet5plus} \\
        Salesforce/codet5p-770m & CodeT5+ & 770M & Multi (9 PLs + En) & \citep{wang2023codet5plus} \\
        uclanlp/plbart-base & PLBART & 140M & Java, Python, En & \citep{ahmad2021unifiedpretrainingprogramunderstanding} \\
        \hline
        \multicolumn{5}{c}{\textit{English Natural Language Models}} \\
        \hline
        facebook/bart-base & BART & 140M & English & \citep{lewis2019bartdenoisingsequencetosequencepretraining} \\
        facebook/bart-large & BART & 400M & English & \citep{lewis2019bartdenoisingsequencetosequencepretraining} \\
        \hline
        \multicolumn{5}{c}{\textit{Greek Models}} \\
        \hline
        dascim/greekbart & BART & Base & Greek & \citep{evdaimon-etal-2024-greekbart} \\
        IMISLab/GreekT5-umt5-base-greeksum & T5 (uMT5) & ~580M & Greek & \citep{giarelis2024greekt5} \\
        IMISLab/GreekT5-umt5-small-greeksum & T5 (uMT5) & ~300M & Greek & \citep{giarelis2024greekt5} \\
        IMISLab/GreekWiki-umt5-base & T5 (uMT5) & ~580M & Greek & \citep{giarelis2024greekt5} \\
        \hline
        \multicolumn{5}{c}{\textit{Polish Models}} \\
        \hline
        allegro/plt5-small & T5 & ~95M & Polish & \citep{chrabrowa2022evaluation} \\
        allegro/plt5-base & T5 & ~275M & Polish & \citep{chrabrowa2022evaluation} \\
        allegro/plt5-large & T5 & ~820M & Polish & \citep{chrabrowa2022evaluation} \\
        sdadas/polish-bart-base & BART & ~100M & Polish & \citep{dadas_polish_bart_base} \\
        \hline
        \multicolumn{5}{c}{\textit{Chinese Models}} \\
        \hline
        OpenMOSS-Team/cpt-large & CPT & 0.4B & Chinese & \citep{shao2021cpt} \\
        uer/bart-base-chinese-cluecorpussmall & BART & 0.1B & Chinese & \citep{zhao2019uer} \\
        uer/bart-large-chinese-cluecorpussmall & BART & 0.4B & Chinese & \citep{zhao2019uer} \\
        OpenMOSS-Team/bart-base-chinese & BART & 0.1B & Chinese & \citep{shao2021cpt} \\
        OpenMOSS-Team/bart-large-chinese & BART & 0.4B & Chinese & \citep{shao2021cpt} \\
        \hline
        \multicolumn{5}{c}{\textit{Dutch Models}} \\
        \hline
        yhavinga/t5-v1.1-base-dutch-cased & T5 v1.1 & 220M & Dutch & \citep{havinga2021t5dutch} \\
        \hline
    \end{tabular}%
    }
    \caption{A list of all embedding models used for the BARTScore evaluation.}
    \label{tab:BART_embedding_models}
\end{table}

\subsection{Results}
\label{sec:neural-results}
\emsejon{We evaluate neural metrics along three dimensions corresponding to our research questions: (i) their alignment with human annotations, (ii) their sensitivity to contextual information, and (iii) their robustness to perturbations. Our goal is to analyze the behavior of these metrics and identify their limitations in multilingual settings.}
\begin{figure}
    \centering
    \includegraphics[width=0.9\linewidth]{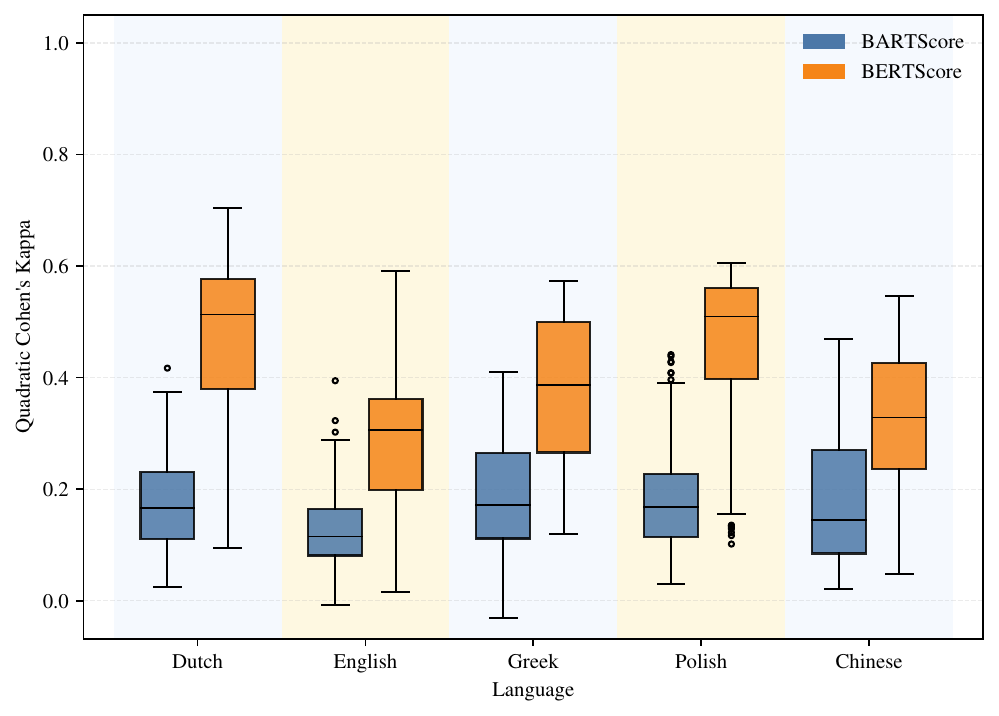}
    \caption{Average Alignment of neural metrics with human experts, divided per language}
    \label{fig:neural-human-alignment}
\end{figure}

\begin{figure}
    \centering
    \includegraphics[width=0.9\linewidth]{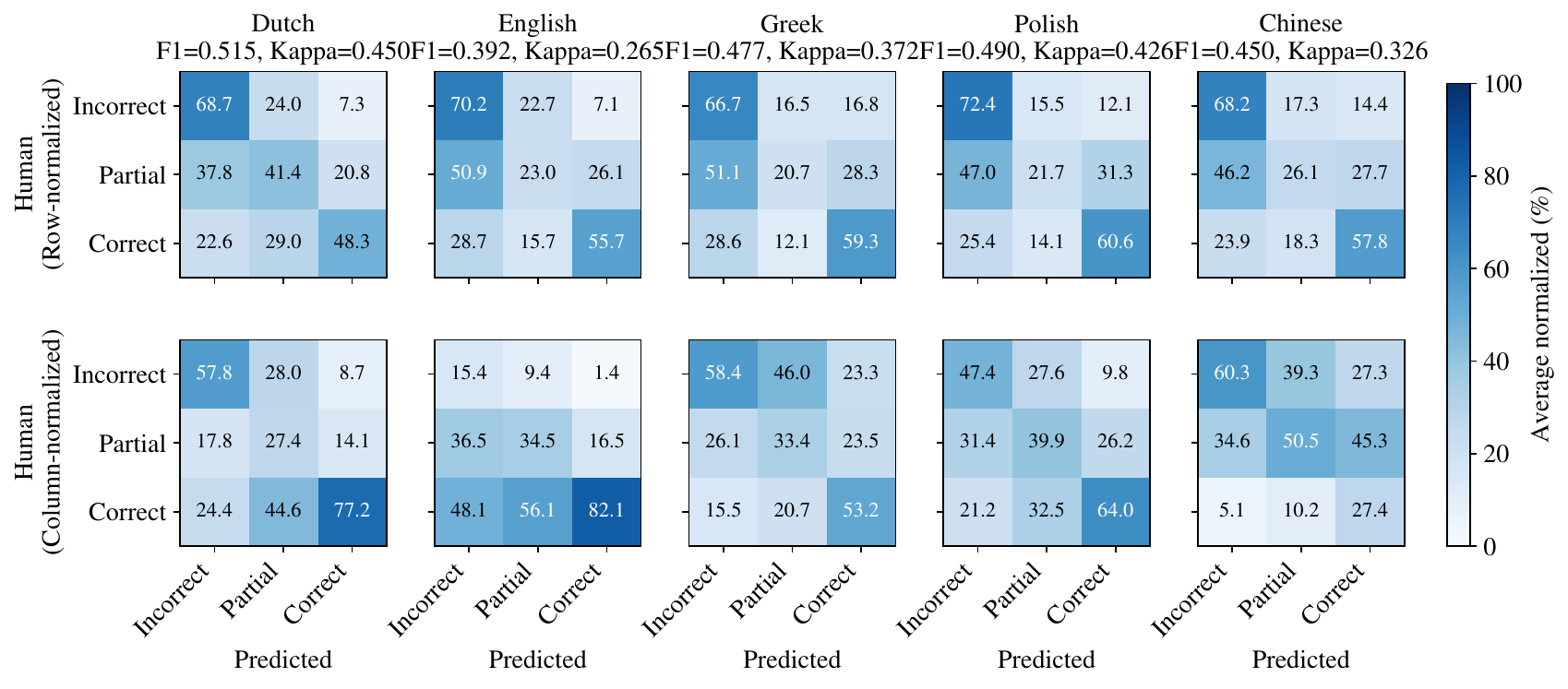}
    \caption{The confusion matrices for the classification of comment generations based on the BERTScore metric, for every language. The top row shows the row normalized matrices, and the bottom row shows the column normalized matrices}
    \label{fig:bertscore-confusion}
\end{figure}

\begin{figure}
    \centering
    \includegraphics[width=0.9\linewidth]{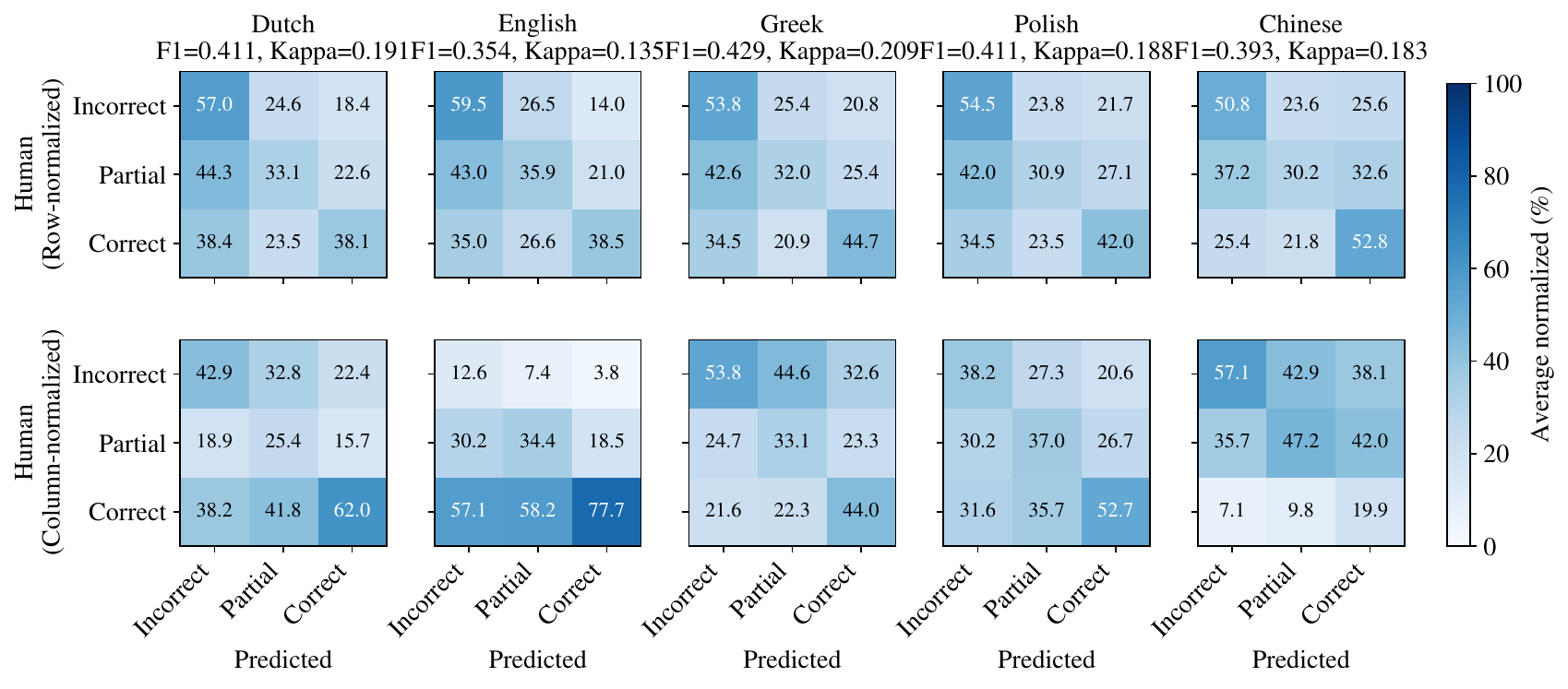}
    \caption{The confusion matrices for the classification of comment generations based on the BARTScore metric, for every language. The top row shows the row normalized matrices, and the bottom row shows the column normalized matrices}
    \label{fig:bartscore-confusion}
\end{figure}

\paragraph{Alignment with human judgment}
\emsejon{Using the large variety of models that we have collected, we first want to know, which models best align with human preference, RQ3.1. We evaluate how well neural metrics align with human annotations by assessing their ability to distinguish between correct, partially correct, and incorrect generations using the classification method described in Section~\ref{sec:Extended_comparison}. We report quadratically weighted Cohen’s kappa between metric-based predictions and human labels. We plot the results for all models combined in Figure~\ref{fig:neural-human-alignment}. Here we see a trend across all languages; BERTScore-based models show higher agreement with human annotations than BARTScore-based models. This indicates that encoder-based metrics more closely align with human judgments of comment quality.}

\emsejon{We also observe a language-dependent trend. Alignment seems to increase for less represented languages, contrary to prior findings in multilingual LLM evaluation where performance typically degrades outside of English. We further investigate this phenomenon and focus on the nature of the misclassifications by plotting the confusion matrices for both the BARTScore and BERTScore classifiers in Figure~\ref{fig:bertscore-confusion} and Figure~\ref{fig:bartscore-confusion}, respectively. Here we plot both the row normalized and column normalized matrices. The row normalized matrices show us the recall of the classifier, while the column normalized matrices show us the precision. Here we see that while non-English languages have more of their mass around the main diagonal, leading to a higher Cohen's Kappa, they tend to be more unpredictable in their errors than English.}

\emsejon{English has the majority of the mass in the lower triangle (especially the lower left corner), this gives it a larger penalty for the Cohen's Kappa score. From a users perspective, however, when focusing on the English BERTScore results, if the generation is predicted as being correct, it would only have a major error in 1.4\% of all cases. While this would be true for 23.3\% of cases in Greek. Furthermore, this trend holds for both BERTScore and BARTScore, and for both Cohen's Kappa and the precision of the \texttt{correct} label; BERTScore outperforms BARTScore.}

\emsejon{Finally, we see that overall, the correct label can be trusted most for all languages except Chinese, scoring the highest recall in all settings except for Chinese. The hardest label to predict correctly was the \texttt{partially correct} label, which had the worst recall and precision for all languages and models, except for English. The \texttt{Incorrect} label had the lowest overall precision in English, yet had a significantly higher precision in all other languages.}

\begin{figure}
    \centering
    \includegraphics[width=0.9\linewidth]{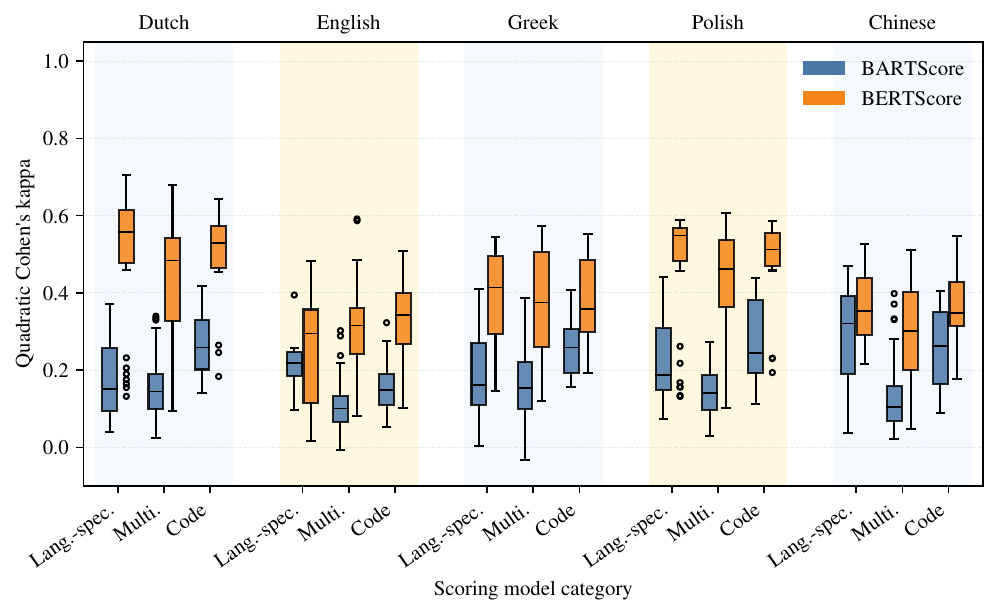}
    \caption{Average QWK of neural metrics with human experts, divided per language, split by scoring model family}
    \label{fig:language-specific}
\end{figure}

\begin{figure}
    \centering
    \includegraphics[width=0.9\linewidth]{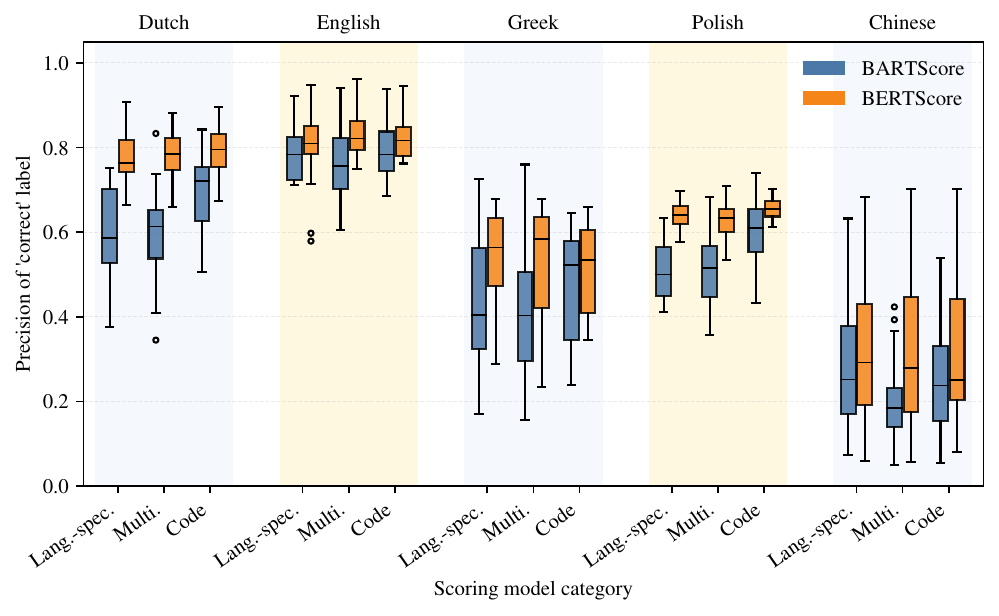}
    \caption{Precision of correct label for neural metrics compared to human experts, divided per language, split by scoring model family}
    \label{fig:language-specifc-precision}
\end{figure}

\subparagraph{Specialized Models}
\emsejon{Having established that performance differs across languages overall, we next examine whether language-specific models help or hinder generation scoring. To do so, we divide the Cohen’s quadratically weighted kappa scores obtained from the BERTScore and BARTScore models into three groups: multilingual models, language-specific models, and code models. Figure~\ref{fig:language-specific} shows the distribution of Cohen’s kappa scores by language, while Figure~\ref{fig:language-specifc-precision} presents the precision for the \texttt{correct} label.}

\emsejon{Across most languages and for both the BERTScore and BARTScore setups, the plots suggest small average differences between these model groups. However, the variance within each group is substantial, making these differences inconsistent. As a result, the findings do not support a simple heuristic for model selection, such as prioritizing language-specific models, in this setting.}

\paragraph{Context}
\emsejon{Next, we answer RQ3.2 and investigate the importance of context for scoring generated comments. As encoder-decoder and encoder-only models depend on the tokens provided as input, incorporating surrounding context should, in principle, enable a more nuanced evaluation of generated outputs, similar to the prompted BARTScore variant proposed in the original work~\citep{yuan2021bartscore}. This is particularly relevant for error types that depend on information from the surrounding code, as without access to this context, such errors are effectively invisible to current neural metrics.}

\begin{figure}
    \centering
    \includegraphics[width=0.9\linewidth]{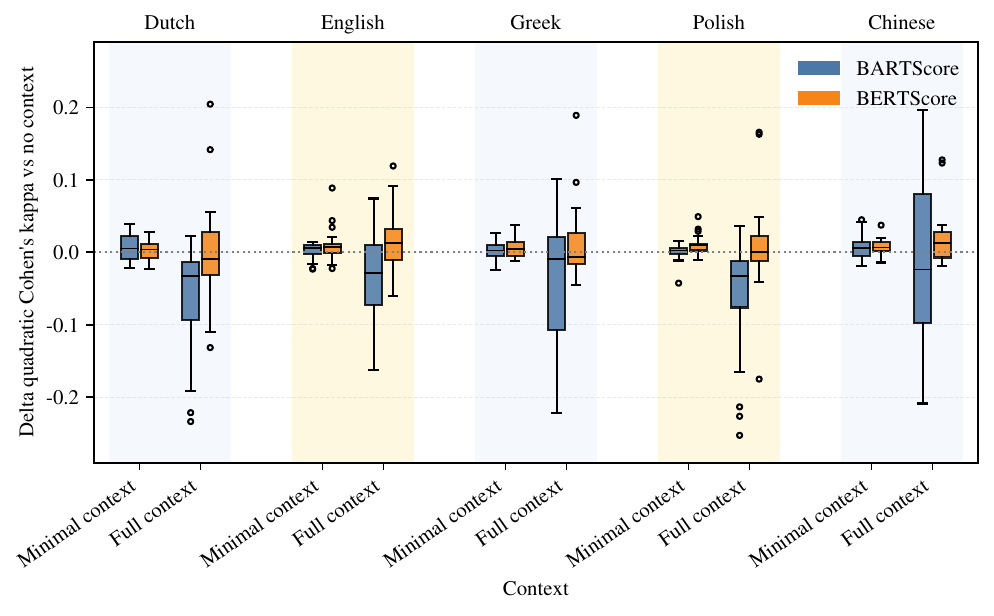}
    \caption{Change in Quadratically weighted kappa compared to human experts, for different amounts of context}
    \label{fig:neural_metric-context}
\end{figure}

\emsejon{To evaluate this, we apply both BERTScore and BARTScore across all samples with varying amounts of context and report the change in Cohen’s kappa relative to the no-context setting in Figure~\ref{fig:neural_metric-context}. The results are mixed. For BARTScore, increasing the context to complete the sentence has little effect on Cohen's kappa, while adding more context leads to a substantial drop in performance across all BARTScore models, in some cases reducing performance to the level of random guessing. In contrast, BERTScore shows greater sensitivity to context. Including minimal context to complete the sentence can significantly improve alignment for some models, and incorporating the full context can further increase performance. However, this effect is not consistent, as for other models performance decreases, resulting in an average change in kappa of $0$ across all languages.}

\emsejon{This high variance in performance change raises questions about the discriminative power of these models across different error types. Since additional context can either improve or degrade performance depending on the model, there is no clear strategy for reliably leveraging context in evaluation. As a result, selecting an appropriate model requires extensive experimentation and human validation to ensure alignment with human judgment for any given task.}

\paragraph{Robustness to noise}
\emsejon{Finally, we answer RQ3.3. We evaluate how robust BERTScore and BARTScore-based models are to noise. To this end, we introduce two types of noise as described in Section~\ref{sec:off-the-shelf_approach} and follow the same methodology used for correctness evaluation in Section~\ref{sec:Extended_comparison}, using a KDE based classifier to assign binary labels indicating noise or no noise. As this is a synthetic binary classification task without human annotations, we report F1 scores instead of Cohen’s kappa.}

\begin{figure}
    \centering
    \includegraphics[width=0.9\linewidth]{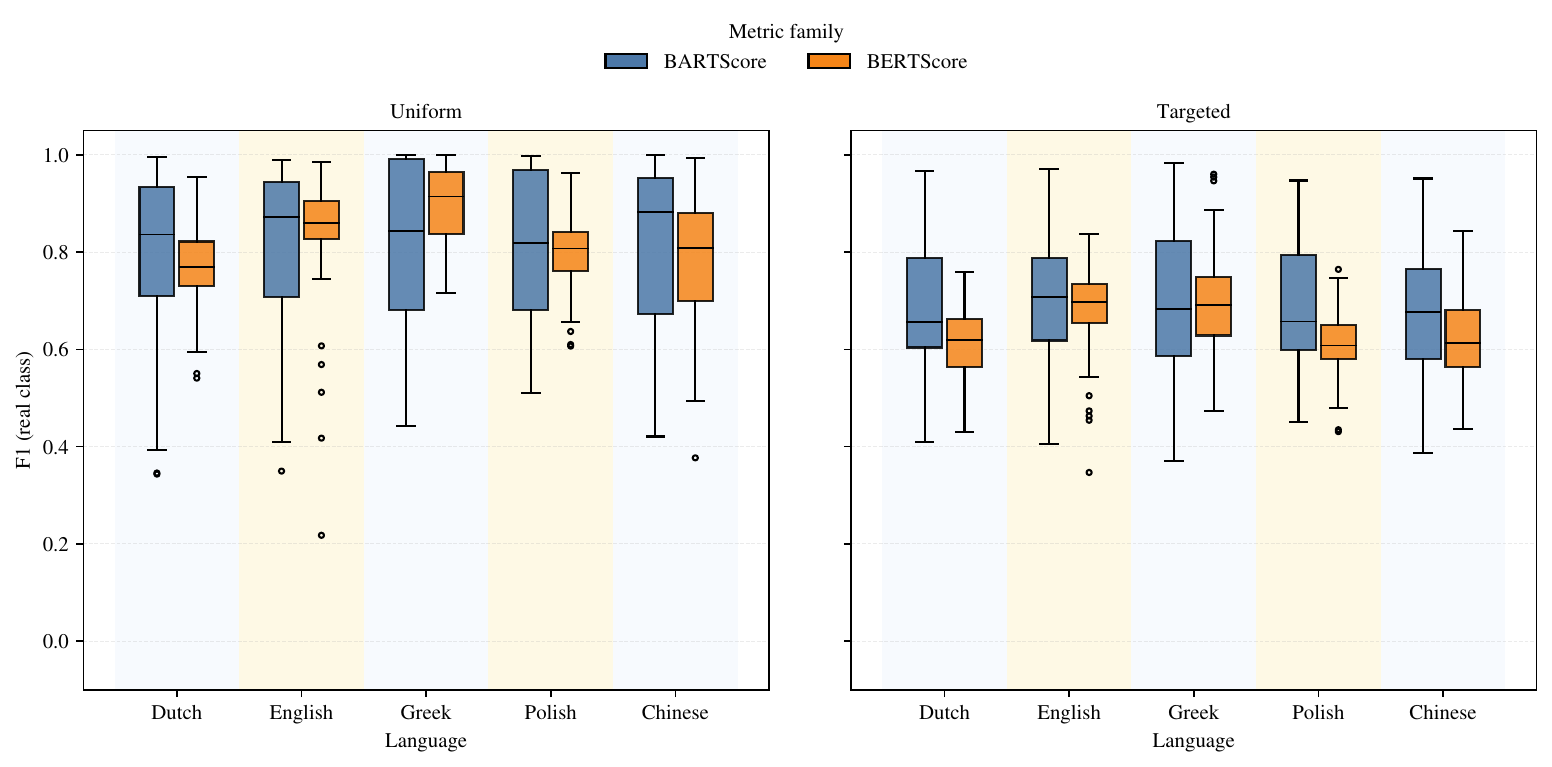}
    \caption{The ability of a BERTScore and BARTScore to separate noise and valid generations.}
    \label{fig:neural_metric-noise}
\end{figure}

\emsejon{Figure~\ref{fig:neural_metric-noise} shows the results across different languages. Contrary to the previous experiments, we see that both BERTScore and BARTScore models perform similarly in these tasks, with BARTScore having a higher variance than BERTScore. We attribute this increase in performance for BARTScore to the similarity of the denoising setup used in training. However, for both metrics, we see that the performance can drop to random guessing, depending on the model.}

\emsejon{When comparing noise types, targeted noise leads to a greater drop in performance than uniform noise. This suggests that the models are scoring generations higher based on the presence of tokens rather than a semantic ordering. This hints that this may also be one of the main causes for the misclassification of accuracy labels in the other experiments.}

\section{LLM-as-a-Judge}
\emsejon{Large language models can also be used to approximate the labeling decisions of human annotators, an evaluation paradigm commonly referred to as \textit{LLM-as-a-judge}. In our final evaluation setting, we adapt existing prompting techniques to approximate expert annotation for multilingual code comment evaluation. Through this setup, we answer RQ4.1, RQ4.2, and RQ4.3.}

\subsection{Approach}
\emsejon{To assess the effectiveness of LLM-as-a-judge in our setting, we implement four prompting techniques that have been proposed to mitigate common shortcomings of LLM-based evaluation. We evaluate these techniques across a representative set of LLMs spanning different providers and model sizes.}

\paragraph{Prompting Techniques}
\label{sec:prompting-techniques}
\emsejon{To ensure a robust evaluation, we assess multiple LLM-as-a-judge strategies for multilingual code comment evaluation. This section describes the shared evaluation setup, the four prompting strategies we implement, their underlying motivation, and how they are instantiated in our pipeline.}

\emsejon{We evaluate four prompting strategies independently: bias mitigation, chain-of-thought reasoning, rubric-based formatting, and hierarchical category evaluation. Each strategy is designed to address a specific limitation of LLM-based evaluation, namely systematic bias, shallow reasoning, ambiguous decision criteria, and cognitive overload when operating over large taxonomies.}

\emsejon{Across all prompting strategies, we use a shared evaluation setup. Each judge receives a prompt consisting of natural-language instructions together with structured JSON components. Specifically, the taxonomy, the original comment, and the five model predictions for a given code snippet are embedded in JSON format as part of the larger prompt. The judge then returns the identified error codes from the taxonomy (Table~\ref{tab:taxonomy}) together with an overall quality label (\texttt{correct}, \texttt{partially correct}, or \texttt{incorrect}), matching the expert accuracy categories defined in Table~\ref{tab:expert_accuracy}. We use a sampling temperature of $0$ with a fixed seed ($42$) to ensure deterministic outputs for reproducibility, and set a maximum output length of $10{,}000$ tokens to prevent infinite repetition. All prompts, including system instructions, evaluation criteria, rubric labels, and output schema field names, are fully translated into the target language of each evaluation set. The response parser maps translated field names back to English keys for uniform processing. For each strategy, we evaluate schema adherence, agreement with human annotators, and error detection accuracy.}


\subparagraph{Bias Mitigation and Standard Prompt}
\label{sec:bias-mitigation}
\emsejon{LLM-based judges are susceptible to several systematic biases that can undermine evaluation validity. The foundational MT-Bench study~\citep{zheng2023judging} identifies three particularly relevant biases in LLM judges: \emph{position bias}, or a preference for responses appearing in certain ordinal positions; \emph{verbosity bias}, or a tendency to favor longer outputs regardless of quality; and \emph{self-enhancement bias}, where judges assign higher ratings to outputs produced by models from the same family. In that study, GPT-4 favored outputs from its own family by approximately 10\%, while Claude-v1 showed an inflation of roughly 25\%. Position bias has since been examined further through systematic analyses of primacy and recency effects in LLM judges~\citep{shi2025judgingjudgessystematicstudy}.}

\emsejon{These biases are particularly relevant in our setting, where the judge evaluates multiple model predictions for the same code snippet simultaneously (Figure~\ref{fig:input_format}). In such a setup, position bias may systematically favor or penalize predictions based on their ordering, while verbosity bias may lead the judge to conflate comment length with comment quality. This is especially problematic in our task, since the expert annotation guidelines (Table~\ref{tab:expert_accuracy}) define correctness in terms of informational completeness rather than length.}

\emsejon{To mitigate these risks, we embed explicit bias-reduction instructions at multiple levels of the prompt. As shown in Figure~\ref{fig:bias-mitigation}, the system message includes three targeted directives: (1) an explicit \textbf{anti-verbosity instruction} (\textit{Do not let comment length influence your judgment of quality. Short comments can be correct; long comments can contain errors.}''), (2) a \textbf{position-independence instruction} (\textit{Evaluate each prediction independently; do not let order influence your judgment.}''), and (3) a \textbf{consistency instruction} (``\textit{Be consistent: apply the same standards regardless of comment length or verbosity.}''). These principles are repeated in the user message to reinforce their application throughout the evaluation.}

\emsejon{The standard baseline prompting setup is limited to these bias-mitigation instructions, a system prompt describing the overall evaluation goal, and a user message containing the step-by-step evaluation instructions. In other words, it does not include any additional reasoning scaffold, rubric formatting, or hierarchical decomposition beyond the core bias-reduction measures described above. This baseline serves as the reference condition against which we compare the more specialized prompting strategies introduced in the following sections. For a full overview of all prompts, we refer to the visualization package provided with the paper.}

\begin{figure}[t]
\centering
\small
\renewcommand{\arraystretch}{1.4}
\begin{tabular}{@{}l@{\qquad}l@{}}
\hline
\textbf{Known LLM Judge Bias} & \textbf{Our Mitigation Strategy} \\
\hline
Verbosity bias & ``Apply the same standards regardless of comment \\
                & length or verbosity'' \\
Position bias & ``Evaluate each prediction independently; ignore order'' \\
Self-enhancement bias & Consistency instruction + independent evaluation \\
\hline
\end{tabular}
\caption{Bias mitigation strategy. Each known LLM judge bias identified in the literature is countered with an explicit prompt-level instruction embedded in both the system and user messages.}
\label{fig:bias-mitigation}
\end{figure}


\subparagraph{Chain-of-Thought (CoT) Reasoning}
\label{sec:cot}
\emsejon{Direct classification prompts can lead to shallow evaluations that lack interpretability and overlook important distinctions, such as the difference between a \emph{partially correct} comment containing a minor variable name error and an \emph{incorrect} hallucinated comment. This distinction is central to our expert accuracy scheme (Table~\ref{tab:expert_accuracy}). G-Eval~\citep{liu-etal-2023-g} showed that requiring LLMs to generate step-by-step reasoning before assigning a score can improve alignment with human judgment, outperforming prior methods on summarization evaluation. The underlying intuition is that explicit intermediate reasoning encourages the model to ground its judgment in observable evidence rather than relying on superficial pattern matching.}

\emsejon{Our CoT implementation guides the judge through seven explicit reasoning steps before classification. These steps are tailored to the structure of our evaluation task. First, the judge must interpret the Fill-in-the-Middle (FIM) masking context (Figure~\ref{fig:input_format}) to identify which code region the comment describes. It must then compare the predicted comment against the original comment and finally apply the inclusion and exclusion criteria defined in our taxonomy (Table~\ref{tab:taxonomy}). Critically, the output schema enforces this reasoning-first procedure: the model must produce a \texttt{reasoning} field containing its step-by-step analysis \emph{before} the \texttt{errors} array, ensuring that justification precedes classification. In addition, each identified error must include a \texttt{confidence} score and a \texttt{justification} field, providing greater transparency into the judge’s decision-making process~\citep{xiong2024can}. We give an overview of the CoT reasoning steps in Table~\ref{tab:cot-steps}. For the full prompt, we refer the reader to the extra material.}

\begin{table}[t]
\centering
\label{tab:cot-steps}
\small
\renewcommand{\arraystretch}{1.35}
\begin{tabular}{@{\,}c@{\;\;}l@{\qquad}l@{}}
\toprule
\textbf{Step} & \textbf{Instruction} & \textbf{Phase} \\
\midrule
1 & Read the predicted comment carefully & \\[-1pt]
2 & Identify the code region via FIM masking context & \textit{Context comprehension} \\[-1pt]
3 & Compare prediction against original comment \& code & \\
\midrule
4 & For each potential error, check \emph{inclusion criteria} & \textit{Criteria verification} \\[-1pt]
5 & Verify \emph{exclusion criteria} before confirming an error & \\
\midrule
6 & Provide reasoning, \emph{then} classification & \textit{Reasoning-first output} \\[-1pt]
7 & Determine overall quality based on the above reasoning & \\
\bottomrule
\end{tabular}
\caption{Chain-of-Thought reasoning steps. The judge follows seven explicit steps, designed around the FIM evaluation context (Figure~\ref{fig:input_format}) and the taxonomy's inclusion/exclusion criteria (Table~\ref{tab:taxonomy}), grouped into three phases: context comprehension, criteria verification, and reasoning-first output.}
\end{table}


\subparagraph{Rubric-Based Taxonomy Formatting}
\label{sec:rubric}
\emsejon{Raw taxonomy definitions, even when expressed in structured formats such as JSON schemas, can remain difficult for LLM judges to operationalize as concrete evaluation criteria. The Prometheus framework showed that structured rubrics with explicit scoring criteria achieve substantially stronger alignment with human evaluators than unstructured instructions, reaching up to 0.897 Pearson correlation~\citep{kim2023prometheus, kim-etal-2024-prometheus}. In addition, \emph{question-specific} rubrics have been shown to outperform generic rubrics in code evaluation tasks by reducing ambiguity in the decision-making process~\citep{Pathak_2025}.}

\emsejon{These findings are directly relevant to our setting. Each of the 26 error categories in our taxonomy (Table~\ref{tab:taxonomy}) was developed through open coding and includes explicit inclusion and exclusion criteria. For example, the Incorrect Synonym (LG-IS) category specifies when a word is “similar but incorrect in context” as an inclusion criterion, and when it should instead be treated as “an acceptable alternative” under the exclusion criterion. To make these definitions more directly usable for LLM judges, we develop an automated rubric formatter that converts the taxonomy into actionable decision templates. As shown in Figure~\ref{fig:rubric-transform}, each JSON-structured error definition is transformed into a rubric entry with explicit \textbf{Mark as PRESENT if} and \textbf{Mark as ABSENT if} sections, directly reflecting the inclusion and exclusion criteria established during taxonomy refinement. This reformulation turns the evaluation task from an open-ended classification problem into a series of guided binary decisions based on the same concrete criteria used by our human annotators, thereby reducing ambiguity and improving consistency. Full rubric examples are provided in the extra material.}

\begin{figure}[t]
\centering
\begin{tikzpicture}[
    node distance=0.3cm,
    jsonbox/.style={
        rectangle, rounded corners=2pt, draw=black!50, fill=white,
        line width=0.5pt, minimum width=4.8cm,
        text width=4.6cm, align=left, font=\footnotesize\ttfamily,
        inner sep=5pt
    },
    rubricbox/.style={
        rectangle, rounded corners=2pt, draw=black!50, fill=gray!6,
        line width=0.5pt, minimum width=5.5cm,
        text width=5.3cm, align=left, font=\footnotesize,
        inner sep=5pt
    },
    heading/.style={font=\footnotesize\bfseries, text=black!70},
]

\node[heading] (jh) {Taxonomy Definition (JSON)};
\node[jsonbox, below=0.2cm of jh, anchor=north, align=left] (json) {%
\texttt{\{} \\
\texttt{\hspace{1.5em}"id": "SE-MD",} \\
\texttt{\hspace{1.5em}"name": "Missing details",} \\
\texttt{\hspace{1.5em}"inclusion": "1) Description} \\
\texttt{\hspace{3em}does not fully describe} \\
\texttt{\hspace{3em}the content. 2) Generated comment is too generic",} \\
\texttt{\hspace{1.5em}"exclusion": "Exclude..."} \\
\texttt{\}}
};

\node[right=0.8cm of json, font=\normalsize, text=black!50] (arr) {$\Longrightarrow$};

\node[heading, right=0.8cm of arr, anchor=west] (rh) {Decision Template (Rubric)};
\node[rubricbox, below=0.2cm of rh, anchor=north] (rubric) {%
\textbf{\#\#\# [SE-MD] Missing details}\\[3pt]
\textbf{Mark as PRESENT if:}\\
\hspace*{0.4em}$\bullet$ Description does not fully\\
\hspace*{0.7em}describe the content\\
\hspace*{0.4em}$\bullet$ Comment is too generic\\[3pt]
\textbf{Mark as ABSENT if:}\\
\hspace*{0.4em}$\bullet$ High-level summary aligns\\
\hspace*{0.7em}with expected response format\\
\hspace*{0.4em}$\bullet$ Comment is general but provides\\
\hspace*{0.7em}sufficient understanding
};

\end{tikzpicture}
\caption{Rubric transformation. The JSON taxonomy definition (left), with inclusion and exclusion criteria from the open-coding process, is automatically converted into an actionable decision template (right) with explicit binary criteria, mirroring the decision process used by human annotators (Table~\ref{tab:taxonomy}).}
\label{fig:rubric-transform}
\end{figure}


\subparagraph{Hierarchical Category Evaluation}
\label{sec:hierarchical}
\emsejon{Presenting all 26 error categories simultaneously introduces substantial cognitive load for LLM judges. Prior work shows that LLM classification performance degrades as the label space grows, while incorporating error relationships and structured label descriptions into the prompt can improve classification accuracy by up to 21\% ~\citep{lee2024improvingllmclassificationlogical}. Providing both error descriptions and their relationships gives the model a clearer decision space and reduces ambiguity during classification.}

\emsejon{Our taxonomy (Table~\ref{tab:taxonomy}) already exhibits a natural hierarchical structure. Its four top-level categories—Model-Specific~(MS), Linguistic~(LG), Semantic~(SE), and Syntax~(ST)—group errors that share evaluation context and decision criteria. We exploit this structure by partitioning the 26 categories into seven semantically coherent clusters, following the conceptual boundaries established during the open-coding process. Specifically, \texttt{linguistic\_grammar} captures surface-level language correctness, \texttt{linguistic\_language} addresses wrong-language issues, \texttt{semantic\_accuracy} and \texttt{semantic\_code} distinguish meaning-related errors from code-snippet errors, \texttt{model\_behavior} groups LLM-specific failure modes such as memorization, repetition, and premature termination, \texttt{syntax\_format} covers comment-formatting issues, and \texttt{meta} contains exclusion and catch-all codes.}

\emsejon{Each cluster is evaluated using a focused prompt that contains only the relevant error categories, together with explicit instructions to \textit{focus only on the error types listed above} and ignore other possible issues. This reduces cross-cluster interference; for example, the judge is not distracted by grammatical issues when evaluating semantic accuracy. The cluster-level outputs are then aggregated in a post-processing step to produce the final per-instance evaluation. This decomposition reduces the effective label space of each evaluation pass from 26 categories to at most 9, placing the task in a range where LLM classifiers have been shown to perform more reliably. The trade-off is computational cost, as this approach requires 7 API calls per instance, one for each cluster. Full cluster prompt examples are provided in the extra material.}

\paragraph{Model selection}
\emsejon{When selecting LLM judges for our setup, we considered several practical and methodological constraints. First, the token requirements for detailed prompts, including the full error taxonomy and corresponding code context, exceed the context windows of models that can feasibly be hosted locally. As a result, we rely on models accessible through APIs~\footnote{openrouter.com}, which introduces significant cost considerations for large-scale evaluation pipelines such as ours.}

\emsejon{Given these constraints, we aim to select a representative set of models. Similar to the neural scoring models, our selection spans multiple dimensions, including flagship models, provider origin (Western vs Chinese), code-specific versus general-purpose models, and variations in model size and architecture (dense vs sparse). While detailed information about the training data of these models is not available, we assume that their scale provides at least some coverage of the target languages. An overview of all selected models is provided in Table~\ref{tab:judges}.}

\begin{table}
    \centering
    \begin{tabular}{c|c|c}
        Model name & Selection Criteria & Citation \\
        \hline
        Claude Haiku 4.5 & Flagship model, large, western & \citep{anthropic_haiku}\\
        Gemini 3 Flash & Flagship model, large, western & \citep{google_gemini3} \\
        Qwen3 Coder Next & Flagship model, code specific, Chinese & \citep{cao2026qwen3} \\
        Qwen3 VL 235B A22B & Sparse, Chinese & \citep{bai2025qwen3} \\
        LongCat Flash Chat & Large, Chinese & \citep{team2025longcat}\\
        Grok 4.1 Fast & Flagship model, Western & \citep{xai2025grok41}\\
        gpt-oss-120b & Mdeium, Western & \citep{agarwal2025gpt}\\
        gpt-oss-20b & Small, Western & \citep{agarwal2025gpt}\\
    \end{tabular}
    \caption{Selected judges, with selection criteria and sources}
    \label{tab:judges}
\end{table}

\subsection{Results}

\paragraph{Technical Challenges}
\emsejon{Before assessing how well LLMs can judge comment quality, we first address RQ4.1 by examining whether LLM judges can reliably produce usable outputs at all. Because our evaluation pipeline depends on structured JSON responses, output-generation failures directly affect the viability of LLM-as-a-judge systems. We therefore analyze two failure modes: parse failures, which make outputs unusable, and empty responses, where models return syntactically valid but uninformative JSON. Together, these issues show that the practical usefulness of LLM judges depends not only on evaluation quality but also on their ability to produce consistent and interpretable outputs.}

\begin{figure}
    \centering
    \includegraphics[width=1\linewidth]{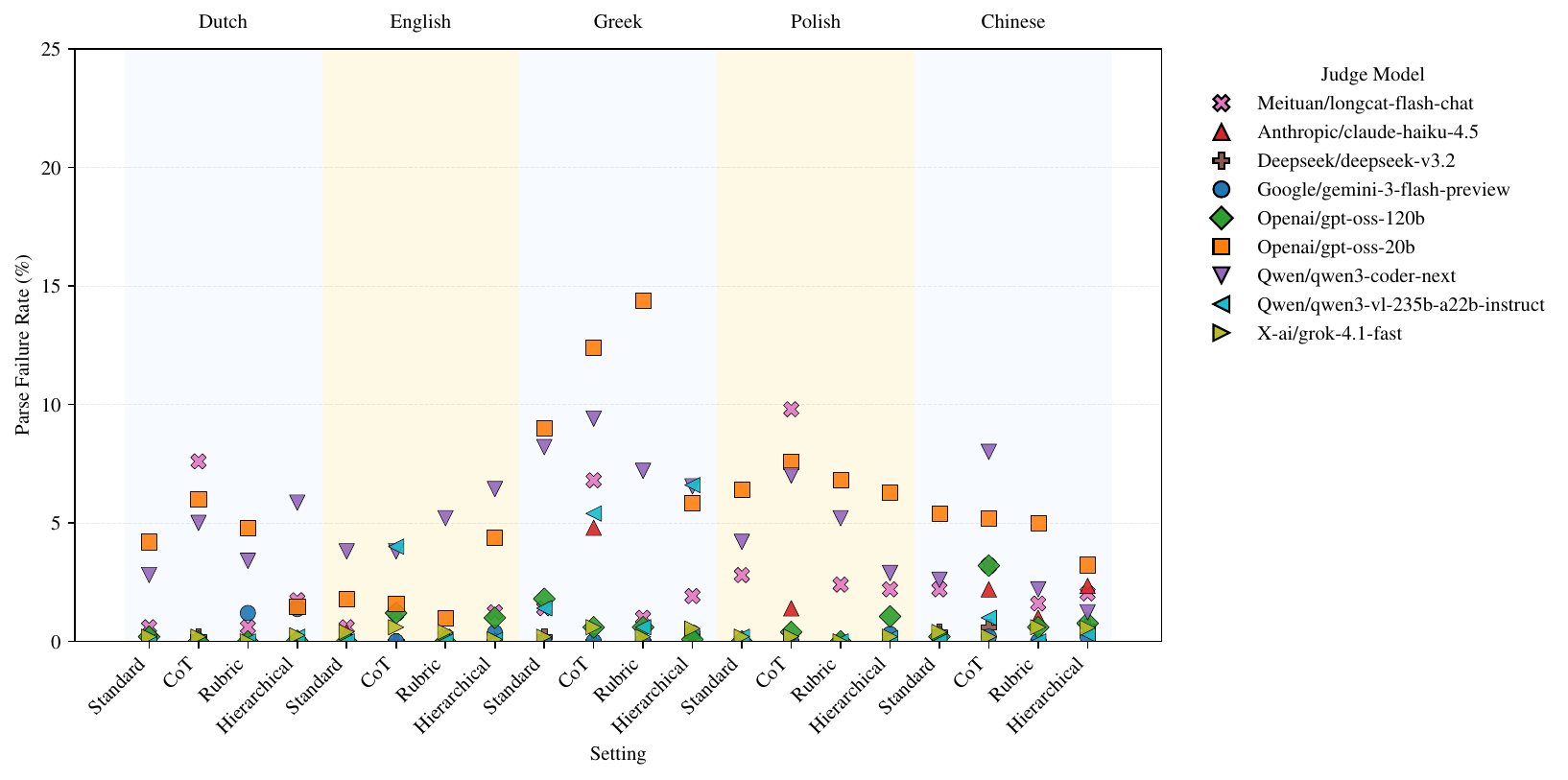}
    \caption{Parse failure rate of LLMs, in all settings being evaluated}
    \label{fig:parse_rate}
\end{figure}

\subparagraph{Parse failure rate}
\emsejon{We first examine whether LLM judges can reliably produce valid outputs by measuring parse failure rates, shown in Figure~\ref{fig:parse_rate}. Failures arise from malformed JSON, infinite repetition that leads to truncated outputs, and over-eager refusals that break the expected schema. Although most models exhibit low failure rates overall, several combinations of model, language, and prompting strategy show substantial degradation. For example, \texttt{gpt-oss-20b} fails on more than $20\%$ of prompts, while \texttt{Meituan\_Longcat} exhibits consistently elevated failure rates under chain-of-thought prompting across all languages. By contrast, \texttt{Qwen3-vl-235b} achieves near-perfect parse reliability in most settings, but still shows a high failure rate for chain-of-thought prompting in Greek. These results demonstrate that the robustness of LLM-as-a-judge pipelines is highly sensitive to both language and prompting strategy, and that output reliability must be established before evaluation performance can be meaningfully interpreted.}

\begin{figure}
    \centering
    \includegraphics[width=1\linewidth]{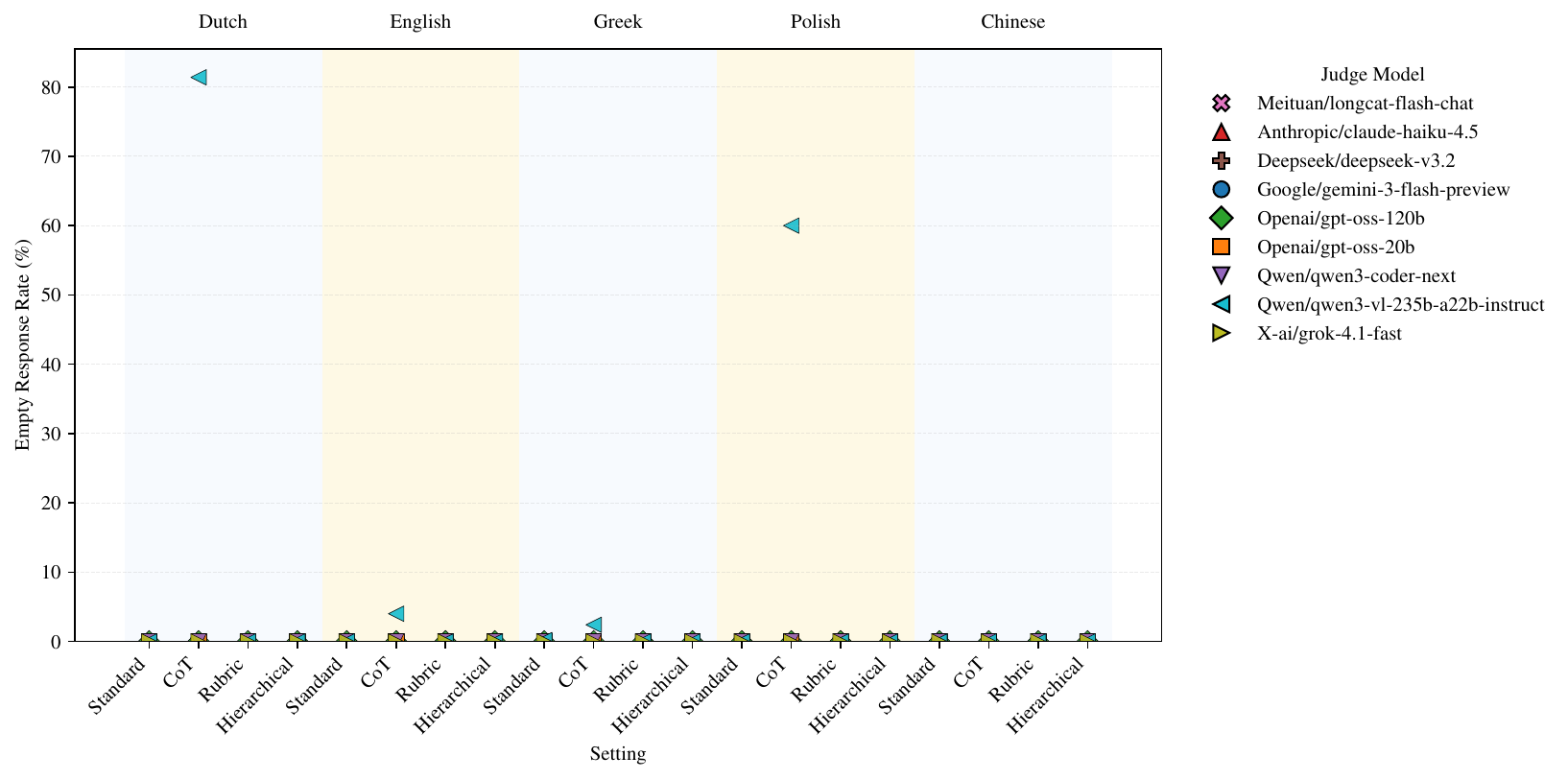}
    \caption{Percentage of outputs that contain correct JSON boilerplate, however the fields themselves are empty}
    \label{fig:empty-json}
\end{figure}

\subparagraph{Empty response rate}
\emsejon{In addition to explicit parse failures, we observe a quieter failure mode in which models return empty JSON outputs, as shown in Figure~\ref{fig:empty-json}. Unlike parse failures, these responses remain syntactically valid but contain no usable information, making them more difficult to detect in practice. Although this behavior is rare overall, it is concentrated in specific model and prompting combinations. In particular, \texttt{qwen3-vl-235b} exhibits severe degradation under chain-of-thought prompting in Polish and Dutch, where more than $60\%$ and $80\%$ of responses, respectively, are empty. In these settings, the judge is effectively unusable despite producing formally valid outputs. As with parse failures, these results show that reliability is not uniform across models, languages, or prompting strategies, but can deteriorate sharply in specific configurations.}

\begin{figure}
    \centering
    \includegraphics[width=1\linewidth]{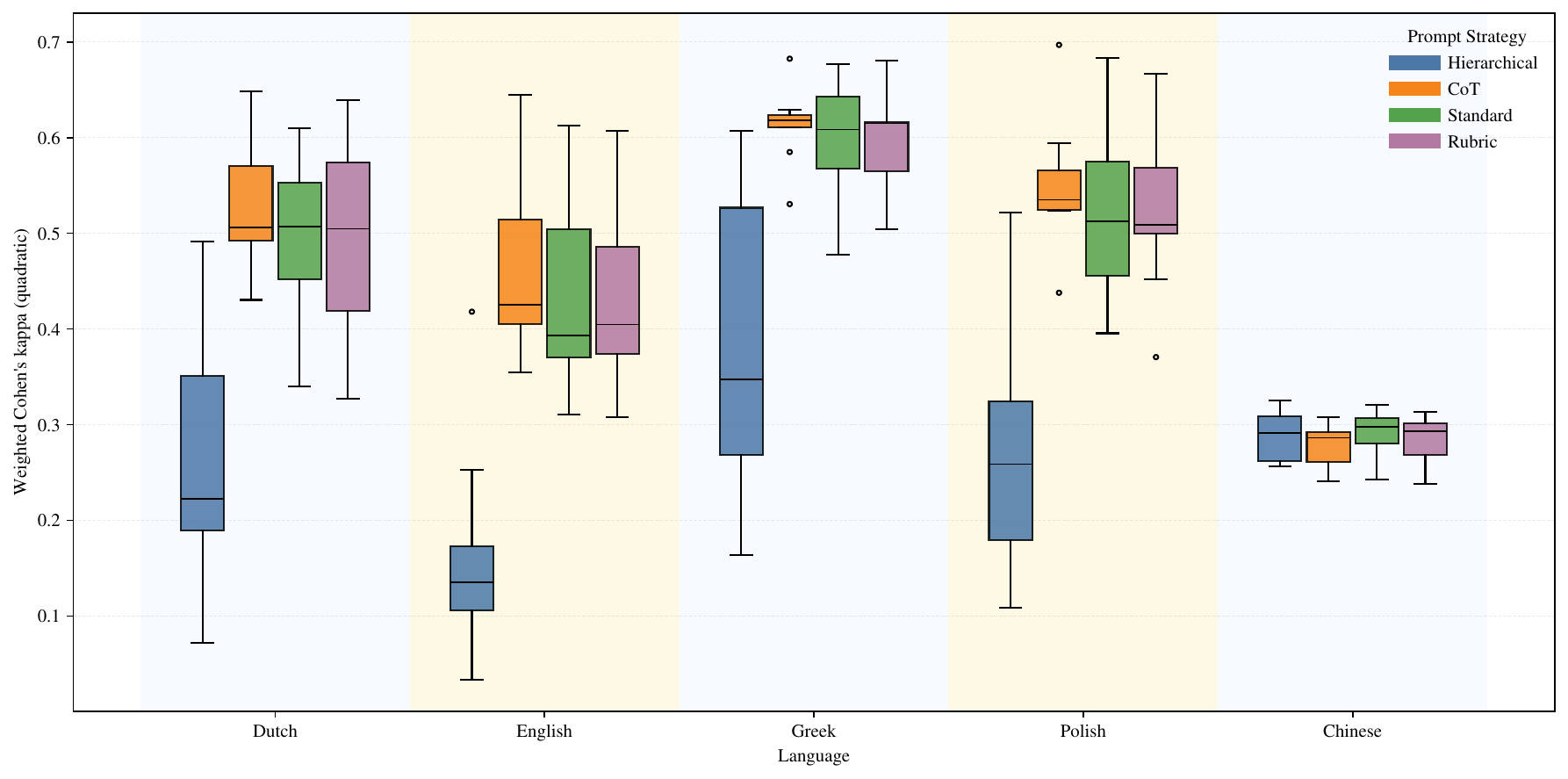}
    \caption{Alignment of LLM-as-a-judge setups with regard to human experts}
    \label{fig:llm-judge-alignment}
\end{figure}

\paragraph{Human Alignment -- Accuracy}
\emsejon{Next, we evaluate the alignment between LLM judges and human experts with respect to the accuracy of generated comments which will answer RQ4.2. We measure agreement using quadratically weighted Cohen’s kappa and aggregate the results by language, as shown in Figure~\ref{fig:llm-judge-alignment}. Similar to the neural metrics evaluation, alignment varies across languages, with English unexpectedly underperforming relative to the other target languages. While this finding runs counter to expectations derived from prior literature, it aligns closely with the language-specific performance differences observed for the neural metrics. Across prompting strategies, chain-of-thought prompting yields the highest agreement, although the differences between prompting strategies are generally smaller than the variance. The hierarchical setting performs substantially worse; however, this can be explained by its rating procedure, which aggregates the outputs of seven separate judge calls, such that an error in any individual call can lower the final rating. For the other settings, it is promising that the best models reach agreement scores in the 0.6--0.8 range, which is commonly interpreted as moderate agreement~\citep{kappa}. In Chinese, however, none of the judges perform well, with all scores remaining just below a kappa value of $0.3$.}

\begin{figure}
    \centering
    \includegraphics[width=1\linewidth]{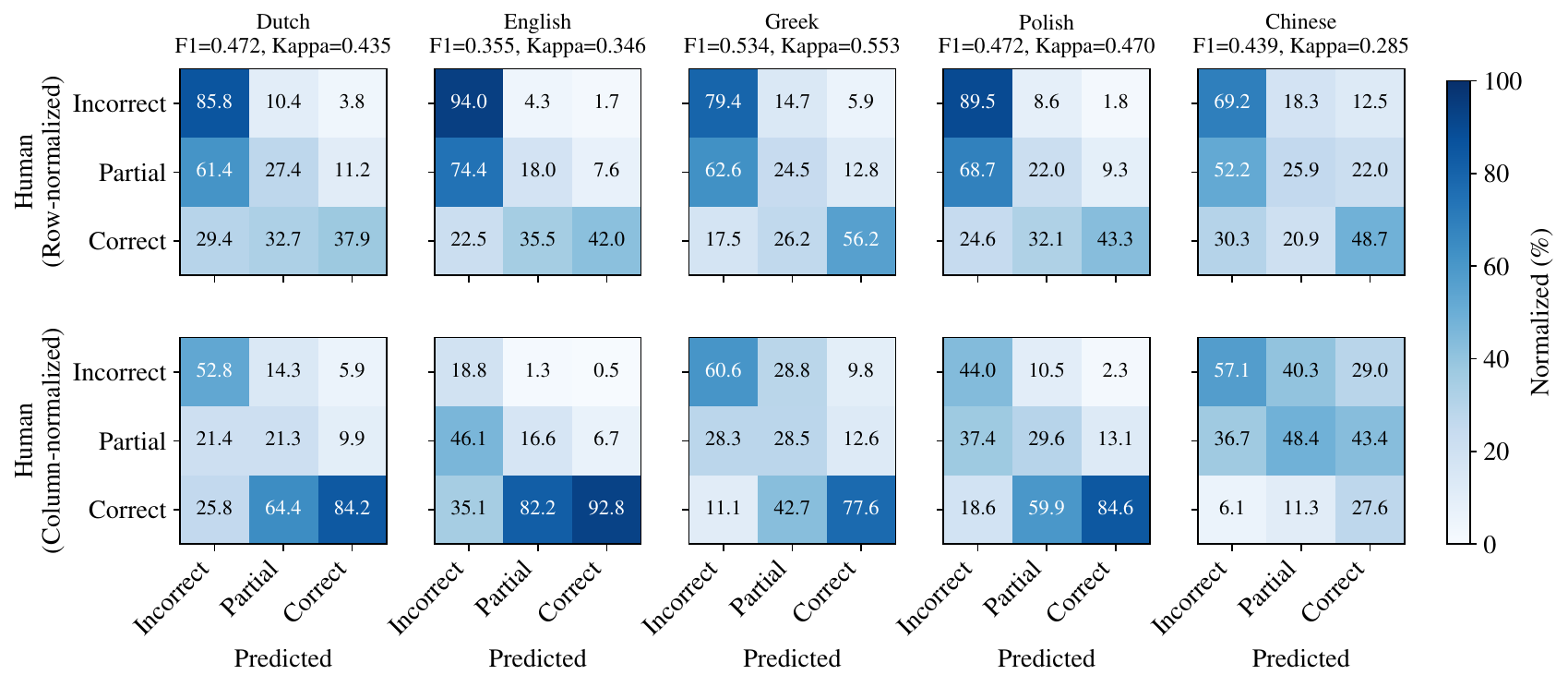}
    \caption{Confusion matrices summarizing agreement between humans and judges. The top row is row-normalized, showing the recall per label, and the bottom row is column-normalized showing the precision of the classifier.}
    \label{fig:llm-judge-cm}
\end{figure}

\emsejon{To better understand the types of errors being made, and to explain why English, against expectations, performs worse according to Cohen’s kappa, we analyze the column- and row-normalized confusion matrices for all languages in Figure~\ref{fig:llm-judge-cm} similar to the procedure used for neural metrics in Section~\ref{sec:neural-results}. In this analysis, we group all prompting strategies together, but exclude the hierarchical setting, as it is penalized more heavily for individual mistakes due to its aggregation setup. The confusion matrices provide three key insights into the alignment between LLM judges and human experts.}

\emsejon{First, LLM judges struggle to capture the nuance of partially correct predictions. This is visible in the middle row of the row-normalized confusion matrices, where predictions labeled as partially correct by human experts are most often classified as incorrect by the judges. This indicates that LLM judges systematically fail to distinguish between \texttt{partially correct} and \texttt{incorrect} outputs.}

\emsejon{Second, when a human expert labels a prediction as \texttt{incorrect}, the LLM judge is also most likely to classify it as \texttt{incorrect}. However, from a user perspective, the more relevant question is the inverse: how often does a human expert agree with the judge’s decision? This is captured by the column-normalized confusion matrices. Here, we observe the opposite trend. When a judge labels a prediction as \texttt{incorrect}, human experts frequently rate the same prediction as at least acceptable. In contrast, when a judge labels a prediction as \texttt{correct}, human experts agree in the vast majority of cases. In English, this agreement is strongest, extending to predictions labeled as partially correct. This implies that LLM judge outputs are only reliably trustworthy when they assign the \texttt{correct} label.}

\emsejon{Finally, examining the overall distribution of mass in the confusion matrices reveals systematic differences across languages. For English, the mass is skewed toward the lower triangle, indicating that LLM judges tend to assign lower ratings than human experts. For other languages, the mass shifts toward the upper-right triangle, meaning that predictions are more likely to be incorrect even when the judge assigns a positive label. This discrepancy in distribution likely explains the lower overall agreement scores observed for English.}

\begin{figure}
    \centering
    \includegraphics[width=0.9\linewidth]{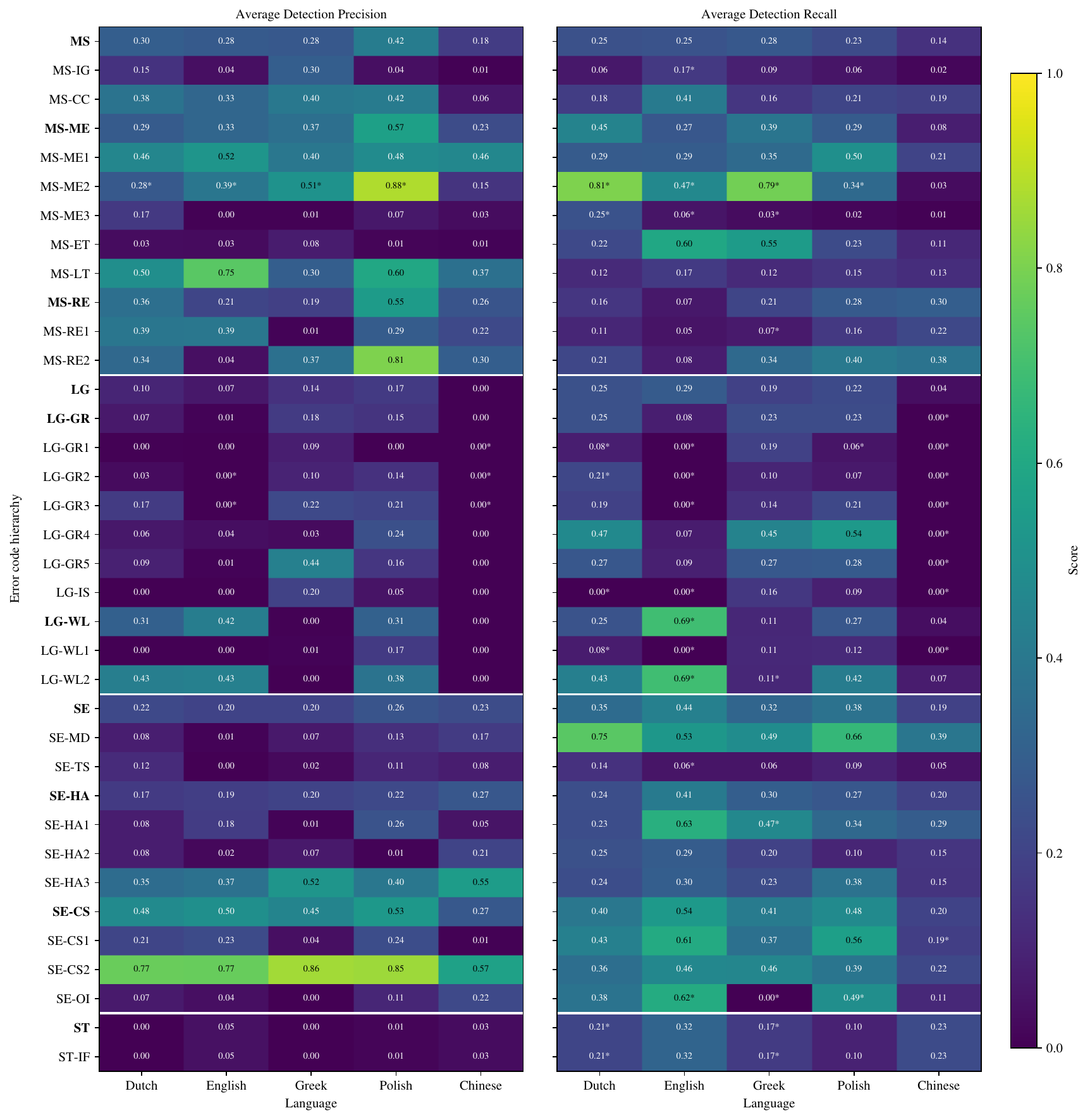}
    \caption{Per error code average precision (left) and recall (right) for LLM judges. Error codes that have been predicted by models less than 10 times, or by human experts less than 10 times are marked with an *}
    \label{fig:judge-errors}
\end{figure}

\paragraph{Human Alignment -- Errors}
\emsejon{Finally, we investigate the performance of the LLM-as-a-judge setups at the level of individual error codes in order to answer RQ4.3. While overall accuracy captures one aspect of alignment, it does not reveal which types of errors LLM judges are actually able to detect. To address this, we report the average precision and recall for each combination of judge model, prompting technique, and error code in Figure~\ref{fig:judge-errors}. In this analysis, we also include the hierarchical setting, since here it is no longer disadvantaged by the aggregation procedure used in the accuracy-based evaluation.}

\emsejon{A striking result is that, for many error labels, average precision is close to $0$. For brevity-related errors such as Early Termination (MS-ET), Missing Detail (SE-MD), and Missing Identifiers (SE-OI), recall is relatively high while precision remains low. This indicates that the judges frequently assign these labels even when human experts do not, suggesting that they systematically overestimate whether comment generations are too generic or lack sufficient detail. A related but inverse pattern appears for Late Termination (MS-LT), where precision is high and recall is very low, indicating that judges are reluctant to assign this label and do so only in a small number of cases that are usually correct.}

\emsejon{In contrast, a small number of more concrete error types achieve relatively strong performance on both precision and recall. This is particularly visible for the inclusion of runnable code (SE-CS2), as well as memorization-related errors involving personally identifiable information (MS-ME1) and URLs (MS-ME2). These results suggest that LLM judges are more reliable when the target error is explicit and easily verifiable from surface-level evidence.}

\emsejon{Looking at the broader pattern, recall is generally higher than precision across most error codes. This suggests that LLM judges are more likely than human experts to assign error labels, reflecting an overall tendency toward over-prediction rather than omission.}

\begin{figure}
    \centering
    \includegraphics[width=0.9\linewidth]{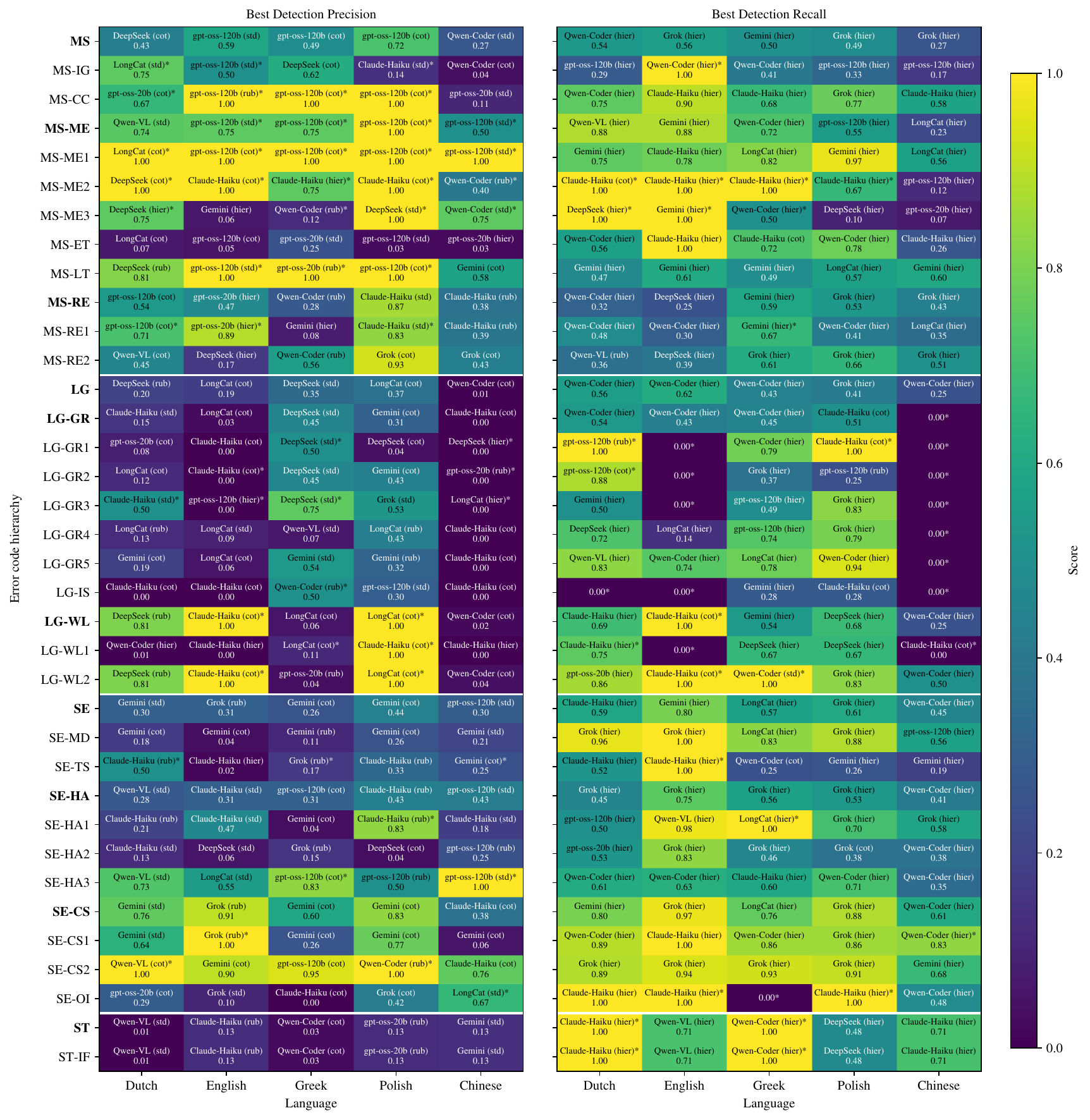}
    \caption{Per error code best performing LLM judge for both precision (left) and recall (right). Error codes that have been predicted by models less than 10 times, or by human experts less than 10 times are marked with an *}
    \label{fig:judge-errors-best}
\end{figure}

\emsejon{To better understand variation across models, we next examine the top-performing setup for each error code, we plot these value if Figure~\ref{fig:judge-errors-best}. This reveals two important findings. First, there is a large discrepancy between the average precision and recall scores and the best observed scores for individual error codes, again indicating substantial variance across models and prompting setups. Second, there is no single model or prompting strategy that consistently dominates in terms of precision: all prompting variants and multiple judge models appear among the top-performing setups for at least one error code. For recall, however, the hierarchical setting achieves the highest scores for most error codes. Given the score distributions, this can largely be explained by the fact that the hierarchical setup returns approximately twice as many error labels as the other settings. In other words, its higher recall appears to come primarily from assigning more labels overall, rather than from a more accurate identification of the errors recognized by human experts.}

\section{Discussion}
\emsejon{This study investigated how the available ML4SE tools support non-English developers in both the generation and evaluation of code comments. Across four complementary evaluation settings, namely expert assessment, off-the-shelf metrics, neural metrics, and LLM-as-a-judge, we observe a consistent pattern: model performance degrades when generating non-English comments, while existing evaluation approaches fail to reliably capture their quality. Off-the-shelf metrics tend to score non-English comments too high, while the types of errors that increase the most outside of English, particularly those related to language quality and semantic correctness, are not reliably detected. In particular, neural metrics are limited by their reliance on the generated comment itself, failing to incorporate information from the surrounding code and thereby missing semantic inconsistencies; even when provided with additional context, their performance does not improve. Similarly, LLM-based judges can leverage the full context, yet still fail to align with human experts when evaluating semantic errors.}

\paragraph{Non-English performance}
The first finding in this investigation was the performance of non-English code generation. We found that depending on the language performance can drop from $81\%$ correct comment generations in English to only $11\%$ in Chinese, for the same model. A drop in performance was to be expected, based on literature~\citep{lai-etal-2023-chatgpt}, however the drop is significantly larger than what is commonly reported. beyond accuracy, we also created a fine grained labeling of failure cases. This showed that the main types of errors that increased in other languages were linked to language erorrs and fluency, as well as semantic errors related to hallucinations. These categories have been shown by previous research to be the key factors limiting adoption of AI models into developer workflow~\citep{hu2022practitioners}. This leads to non-English developers being disadvantaged far more for their preferred language than they are in other LLM systems. This further demonstrates that there is a large gap between the support of different users that needs to be addressed, and has been largely invisible in scientific literature up to now.

We attribute this discrepancy in performance to non-English code being disproportionately resource scarce compared to other non-English languages online, due to developers being discouraged from contributing to open source projects in other languages~\citep{bhuiyan2026write}, and recommendations to exclude non-English from LLM training procedures~\citep{vitale2024catalog}. 

\paragraph{Alignment}
\emsejon{The first quality of a good metric we will address is alignment with humans.
For both the neural metric and the LLM-as-a-judge setting, we saw the common trend that the metrics are overly critical in English; however, they have a high precision for the \texttt{correct} label and are less critical for other languages, being more likely to score a comment generation as better than it is. As a result, this will make the non-English performance of code LLMs appear to be better than they are in reality, obscuring the gap in performance.}

\emsejon{More practically, automatic scores are often used to rank models and guide deployment decisions. If a metric is more likely to score incorrect non-English generations as correct, it introduces a systematic bias into model selection. Furthermore, synthetic data is more commonly used in deep learning pipelines, where overestimating the ability of a data generation model will result in lower quality data being introduced into the training process, compounding the discrepancy. This makes is essential to recognize this systematic bias in the evaluated metrics, and to address it in future research.}

\paragraph{Error Coverage}
\emsejon{Next, we investigated what types of errors the metrics were able to detect. This was done by adding context to the neural metrics, and using the taxonomy from the open coding section to generate error codes per comment.}

\emsejon{Our open-coding investigation showed that the number of errors related to the factuality of the comments increased by at most a factor of $3$ compared to English, while the number of errors related to language increased by a factor of $15.1$. This larger decrease in the models' ability to generate accurate natural language damages the trust users have in the model and is one of the main requirements practitioners name for automatic comment generation~\citep{hu2022practitioners}. Furthermore, the level of detail (SE-MD) drops significantly (up to $21.8\times$ for Chinese), which is another factor limiting the adoption of models by practitioners.}

\emsejon{For the neural metrics, we see that adding context generally worsens human alignment for BARTScore, and is inconclusive, improving and decreasing alignment depending on the model for BERTScore. The consequence of this is that we see no evidence that neural metrics can be relied upon for detecting errors that are reliant on the context. Hallucinations, required level of detail, copying of context, and Omitted identifiers are all errors that were identified in the open coding investigation; however, they cannot reliably be detected with neural metrics.}

\emsejon{From the LLM-as-a-Judge setting, the Judges always had access to the full context for each comment.
Yet, they performed poorly in the categories related to Early termination, specificity of the comment, and had only weak agreement with human experts on in-context hallucinations. Furthermore, the judges had a very low average precision for language errors.}

\emsejon{The main implication of these findings is that current evaluation methods remain weakest on the categories that matter most for practical adoption. Neural metrics do not reliably incorporate the full code context, which makes context-dependent issues such as hallucinations, insufficient detail, copied context, and omitted identifiers difficult to detect. LLM-as-a-judge setups perform better on more surface-level problems, such as explicit inclusions of code, but still fail to reliably identify several of the categories that developers consider most harmful in practice~\citep{hu2022practitioners}. This limitation is particularly concerning because these same categories increase substantially in non-English code comment generation, meaning that current evaluators are least reliable precisely where the underlying generation problem becomes most severe.}

\paragraph{Robustness}
\emsejon{Next, we look at the robustness to noise of the neural metrics. From the research we have conducted in this investigation, we see that not all metrics can capture all types of errors. However, for a metric to be considered trustworthy, it should be able to detect the difference between a genuine prediction and random noise.}

We showed that from an overall score point of view, this was not the case for off the shelf neural metrics, where noise had a significant overlap with the scores assigned to genuine predictions. This raises questions about what the metrics actually score. \emsejon{We extended these results to see if it was a consistent issue across any model being used as the basis for a neural metric, and we show that it is very inconsistent. For both BARTScore and BERTScore metrics, some models had close to perfect separation, while others were worse than random. Consequently, the fact that uniform noise was easier to separate than targeted noise shows that the neural metrics are still using some element of token matching when generating their scores. This inconsistency between models would require each base model to be evaluated separately before being used, as noise should be separable from a genuine prediction for a reliable metric.}

\emsejon{To increase trust in the outputs of neural metrics and, in turn, their usability, these are both criteria that should be taken into consideration. As the scores are based on trained models, we see potential in changing the training procedure in such a way as to give more trust that the output of the metric will not be vulnerable to tokens in the context and will be able to reliably separate noise.}

\paragraph{Sensitivity}
\emsejon{Finally, we focus only on the usability of the LLM-as-a-judge setup. While it performed better than the neural metrics for scoring the generations, its sensitivity to specific settings needs to be addressed. We saw that some models failed to produce usable output in up to 80\% of all cases regarding empty outputs, or around 15\% of prompts where the JSON was malformed.}

\emsejon{Importantly, there was no clear correlation between languages and prompt settings with failure rates. The same models that were returning empty json in 80\% of all responses had no issues when switching prompting strategy. Furthermore, the parse failures can also vary by 15\% between different prompts within a language. The only general trend we notice is that, overall, in all models and prompting settings, non-English prompts seemed to have a slightly higher failure rate.}

\emsejon{The biggest issue with these findings for the adoption of LLMs as judges is that it shows that they are unpredictable. This means that users will need to run tests to validate that the Judges will return valid responses for their use case, and if extending their usage, it can happen that users will need to switch models as the new setting (prompting, or language) can cause the model to break. All of this adds extra costs to the adoption of the LLM judges into any real world pipelines.}

\paragraph{Practical Recommendation}
\emsejon{Based on the findings from all four investigations, we derive the following
practical recommendations for working with non-English code:}

\begin{enumerate}
    \item \emsejon{\textbf{Evaluate models per language prior to deployment.}
    Model performance varies substantially across languages, and improvements
    observed in English do not reliably transfer to other languages. For example,
    when comparing CodeQwen~1.5 and CodeGemma, English performance suggests an
    expected improvement of 27.6\% in correct predictions, while in Chinese the
    same change leads to a decrease in performance. This highlights the need to
    validate model behavior separately for each target language.}

    \item \emsejon{\textbf{Interpret metric predictions asymmetrically.}
    Across most languages, evaluation metrics are more reliable when predicting
    that a generation is \texttt{correct} than when distinguishing between
    \texttt{partially correct} and \texttt{incorrect} outputs. This asymmetry
    suggests that metrics are better suited for confirming high-quality
    generations than for detecting and categorizing errors. However, this
    behavior does not hold consistently across all languages; in Chinese, none
    of the evaluated metrics demonstrate reliable agreement with human annotations.}

    \item \emsejon{\textbf{Prefer BERTScore over BARTScore, rely on relative rankings, and account for high model variability.}
    Neural metrics such as BERTScore and BARTScore provide a computationally
    efficient approach to evaluating generated comments, but they are limited
    in their ability to distinguish between different levels of generation
    quality compared to LLM-as-a-judge setups. Performance varies substantially
    across scoring models, making model selection a critical factor. Heuristics
    such as selecting models based on target language or code specialization do
    not reliably lead to better performance and can be misleading in practice,
    as alignment varies significantly even within these categories.}

    \emsejon{Across all settings, BERTScore metrics consistently outperform BARTScore
    metrics. However, absolute metric scores do not reliably reflect generation
    quality and can even suggest the opposite ranking of models compared to
    human judgment. As a result, relative rankings between generations are
    essential for meaningful comparison, and should not be used between languages.}

    \item \emsejon{\textbf{Use LLM-as-a-judge pipelines cautiously due to cost, instability, and deployment risks.}
    LLM-based evaluators achieve the highest overall agreement with human
    annotations and provide more fine-grained assessments than neural metrics,
    but they come at substantially higher computational cost. In addition,
    their behavior can be unpredictable. Depending on the combination of language, model, and prompting strategy, it can result in models not outputting usable responses, which cannot be effectively predicted.}
    
    \emsejon{In a deployment setting, this creates a practical limitation: extending a
    pipeline to new languages may require reconfiguring the entire evaluation
    setup, including the choice of model and prompting strategy, in order to
    maintain consistent behavior. As observed in our experiments, the most
    important factor for alignment is the choice of model rather than the
    prompting strategy, and this choice also determines whether the judge
    returns a usable output.}

    \emsejon{Despite their strong alignment, LLM judges fail to reliably detect errors
    related to language fluency and semantic correctness, which are among the
    most important factors for adoption by practitioners. These limitations make
    LLM-as-a-judge pipelines unsuitable as standalone replacements for human
    evaluation, particularly in multilingual scenarios.}
\end{enumerate}

\section{Threats to Validity}
\emsejon{Finally, to assess the validity of our method, we structure the threats to validity according to the standard framework used in empirical research: internal validity, external validity, conclusion validity, and construct validity~\citep{threats}.}

\subsection{Conclusion Validity}
\paragraph{Labeling Bias}
This study is subject to the risk of labeling bias, as only a limited number of experts labeled errors for each language. We mitigate this risk by defining clear inclusion criteria for each error label and by conducting iterative discussions among the six authors to harmonize these classifications. Nevertheless, labeling bias remains a concern in a multilingual setting despite these mitigation efforts. To support transparency and independent verification, we release all labeled data publicly.

\subsection{Internal Validity}
\paragraph{Memorization Risk}
Most of the models investigated in this study do not disclose their training data. As a result, we cannot determine whether memorization influenced model outputs on the tasks under evaluation. Because non-disclosure of training data is common, we are unable to control for this threat directly. In cases where memorization was particularly obvious, we annotated it as a separate error category. We encourage open-source model releases to include training data documentation so that future evaluations can be conducted more fairly and transparently.

\paragraph{Judge Bias}
\emsejon{Although we include a broad set of leading models across multiple categories, only a small number of providers currently have the resources to develop models at the scale required for this investigation. This creates a risk of family-level judge bias. For example, comments generated by CodeGemma may be evaluated by related models such as T5Gemma or Gemini. Prior work suggests that, in pairwise evaluation settings, models may prefer outputs produced by similar model families~\citep{li2025preference}. However, our setup differs from a direct pairwise preference task, and the large variety of generation and evaluation model combinations reduces the likelihood that such effects systematically determine the overall findings. Even so, this bias cannot be ruled out entirely.}

\subsection{Construct Validity}
\paragraph{Limited Model Size}
This investigation focuses on comment-generation models in the 7B--8B parameter range. Although larger models often produce stronger outputs, we restricted model size to the largest range that was consistently available across all architectures under investigation. This reduces the risk of introducing scale as a confounding factor and allows for a fairer comparison between model families.

\emsejon{This threat primarily affects the results for RQ1, which concerns comment generation quality. For the remaining experiments, the generated comments serve mainly as evaluation inputs, meaning that the precise model size used for generation is less central to the validity of the results. As a result, this threat is largely limited to the first research question.}

\paragraph{Data Source}
\emsejon{For this investigation, we collected data by searching GitHub using keywords in multiple languages. This sampling strategy introduces bias, as the resulting dataset necessarily contains at least one matched search term and may therefore over-represent certain types of files. We mitigate this effect by limiting the number of files contributed by each keyword. In addition, sourcing files directly from public repositories may introduce variation in code quality, which may in turn affect LLM generation performance. Unfortunately, there is currently no established source of high-quality non-English code that could be used instead, and constructing such a dataset is beyond the scope of this study.}

\paragraph{Optimization}
\emsejon{The goal of this investigation was to provide a broad overview of the challenges and opportunities of ML4SE tools for non-English developers. As a result, our study design spans a diverse set of models, languages, and evaluation settings, making exhaustive optimization for each individual configuration prohibitively expensive. It is therefore likely that alternative hyperparameter choices or different prompt phrasings could improve performance in specific settings or languages. However, our conclusions are based primarily on the general trends observed across experiments rather than on absolute peak performance in any single configuration. In addition, we follow the best practices established in prior work in good faith when selecting prompts and experimental settings. For this reason, although targeted optimization may shift individual results, we do not expect such changes to be large enough to alter the overall conclusions of the study.}

\subsection{External Validity}
\paragraph{Judge Model APIs}
\emsejon{The scale of the state-of-the-art models used in the LLM-as-a-judge setting makes local hosting infeasible for most academic institutions. This forces reliance on third-party API providers, which may apply undocumented preprocessing steps or inject additional system instructions into the input. Reproducibility also depends on trusting that the API correctly implements features such as fixed random seeds. Finally, because these models are offered through commercial providers, access may change over time: providers may discontinue a model, change its implementation, or cease operations altogether. This creates an additional threat to the long-term reproducibility of the study.}

\section{Future Work}
Future work should begin with a closer investigation of the multilingual nature of code itself. Although the community generally recognizes that code can be written in non-English languages, the extent to which this occurs at scale is still largely unknown. There is also limited research on how non-English language is distributed within code, whether in comments, function names, variable names, or strings, and how these elements interact with English within the same artifact. This challenge was already central to the construction of our dataset. Although automatic language detection tools exist~\citep{ercdidip2022}, they struggle in code settings because of code-switching~\citep{ojo2025divers}, transliteration~\citep{10.1145/2872518.2890560}, and mismatches between training data and real-world usage~\citep{ojo2025divers}. These difficulties are even more pronounced in code, where natural language is interleaved with programming syntax, identifiers do not follow standard grammatical patterns, and English keywords dominate most programming languages. Edge cases, such as Greek symbols in mathematical contexts, can also lead to false positives. As a result, reliably identifying and collecting non-English code remains difficult. This lack of data is likely reflected in the reduced performance of the initial LLMs used to generate comments, as we observe lower performance across all languages. Being able to effectively locate non-English code would be the first step in supporting non-English developers further.

\emsejon{A second important direction for future work is the development of stronger evaluation methods. In our investigation, neural methods often struggled to distinguish between correct and incorrect predictions. In addition, comparisons based only on average scores could be misleading, particularly when random noise was introduced. These findings resemble results from other work on multilingual embeddings of code tokens~\citep{katzy2023impact}, where cosine similarity between code tokens increased depending on the number of training steps. Building on this, future work could use our human-annotated dataset as a benchmark for measuring how well neural metrics separate different types of errors. Such an accessible alignment test could support more targeted pretraining, including noise-aware or contrastive learning approaches~\citep{Clark2020ELECTRA:}, as well as negative sampling~\citep{mikolov2013efficientestimationwordrepresentations}, to increase the distance between embeddings that represent meaningfully different outputs while minimizing the distance between aligned ones. Our taxonomy of errors could also serve as a basis for generating more realistic synthetic negative samples, instead of relying on random corruptions~\citep{lewis2019bart}. The broader goal of this work should be to develop evaluation models that can be used consistently for code artifacts, remain robust to noise, align with human judgment, and generalize across languages.
}

Finally, recent research on using LLMs to replace manual annotation of software engineering artifacts~\citep{ahmed2024llmsreplacemanualannotation} suggests that, although LLM-generated annotations can align with human annotations, they remain costly to evaluate. To support further progress in this area, we invite the community to use our publicly available dataset to compare model outputs against expert annotations. This can help advance the development and evaluation of automated annotation systems while grounding their performance in human judgment.

\section{Conclusion}
We explored the support for non-English developers in the emerging ML4SE tool ecosystem. We focused on the ability of 5 LLMs to generate comments in 4 non-English languages (Chinese, Dutch, Greek, Polish), as well as a baseline of English comments to compare to. First, we conducted an open coding study, where experts fluent in each target language rated the generated comments as either \texttt{correct}, \texttt{partially correct}, or \texttt{incorrect}. This showed that there was a definite decrease in the target LLMs ability to generate comments in all target languages other than English. Along with the accuracy of the comments, we also curated a taxonomy of common failure cases across all languages. Next, we evaluated the generated comments using available out of the box metrics commonly used to score LLM outputs. We showed that neural metrics do not generalize well across languages. Languages that scored low based on expert evaluations scored higher than English when using neural metrics. Furthermore, we investigated how robust neural metrics are to noise and showed that there is a significant overlap between the scores real LLM outputs received and synthetic noise. \emsejon{Following this investigation, we evaluated how non-English developers could be supported by neural metrics and extended the existing methods by using more modern models, language specific models, and code specific models. We showed that using specialized models only marginally increases alignment with human experts, averaged across all models, and the best performing models are still multilingual models in most cases. Furthermore, we showed that for all languages, the ability of neural models to separate noise from real generations varies widely from almost perfect to worse than random guessing. Finally, we showed that using LLM-as-a-judge is the best solution for scoring the accuracy of LLM outputs, scoring higher in alignment on 3 out of 4 prompting techniques than BERTScore or BARTScore. This, however, comes at a far higher cost for inference, as well as LLM specific shortcomings such as not outputting parseable JSON, false refusals, or empty responses, which can occur in up to 80\% of all outputs depending on language and prompting template. Furthermore, the LLM judges agreement with human experts varied largely between error-codes, and between models, showing the importance of a high quality evaluation dataset.}

\bibliography{sn-bibliography}

\end{document}